\documentstyle[12pt,pstricks,psfig]{article}
\newcommand{\bold}[1]{\mbox{\boldmath $#1$}}    
\newcommand{\ie}{{\em i.e.}}                    
\newcommand{\etal}{{\em et al.}}                
\newcommand{\etc}{{\em etc.}}                   
\newcommand{\MeV}{{\rm MeV}}                    
\newcommand{\fm}{{\rm fm}}                      
\newcommand{\fbar}{\bar{f}}                     
\newcommand{\rme}{{\rm e}}                      
\newcommand{\r}{{\bold{r}}}                     
\newcommand{\p}{{\bold{p}}}                     
\newcommand{\k}{{\bf k}}                        
\newcommand{\beq}{\begin{equation}}
\newcommand{\eeq}{\end{equation}}
\newcommand{\beqar}{\begin{eqnarray}}
\newcommand{\eeqar}{\end{eqnarray}}
\newcommand{\bfig}{\begin{figure}}
\newcommand{\efig}{\end{figure}}
\newcommand{\ul}[1]{\underline{#1}}             
\newcommand{\del}{\partial}                     
\newcommand{\eps}{\epsilon}                     
\newcommand{\pphi}{{\bold{\phi}}}               
\newcommand{\ppsi}{{\bold{\psi}}}               
\newcommand{\pphit}{{\bold{\phi_\perp}}}        
\newcommand{\dpphi}{{\delta\bold{\phi}}}        
\newcommand{\dppsi}{{\delta\bold{\psi}}}        
\newcommand{\dphi}{\delta\phi}                  
\newcommand{\dphip}{\delta\phi_\parallel}
\newcommand{\dphit}{\delta\phi_\perp}
\newcommand{\mup}{\mu_\parallel}
\newcommand{\mut}{\mu_\perp}

\begin{document}
\begin{titlepage}

{\sl Physical Review D}\hfill LBL-38125\\[8ex]

\begin{center}
{\large {\bf
Statistical Properties of the Linear Sigma Model$^*$}}\\[8ex]
{\sl J\o rgen Randrup}\\[1ex]
Nuclear Science Division, Lawrence Berkeley National Laboratory\\
University of California, Berkeley, California 94720
\\[6ex]

January 12, 1996\\[6ex]
{\sl Abstract:}
\\
\end{center}

{\small\noindent
The statistical equilibrium properties of the linear sigma model are studied,
with a view towards characterizing the field configurations
employed as initial conditions for numerical simulations
of the formation of disoriented chiral condensates
in high-energy nuclear collisions.
The field is decomposed into its spatial average, the order parameter,
and the fluctuations, the quasi-particles,
and enclosed in a rectangular box with periodic boundary conditions.
The quantized quasi-particle modes are described approximately by
Klein-Gordon dispersion relations containing an effective mass
that depends on both the temperature and the magnitude of the order parameter.
The thermal fluctuations are instrumental in shaping the effective potential
governing the order parameter,
and the evolution of its statistical distribution with temperature is discussed,
as is the behavior of the associated effective masses.
As the system is cooled the field fluctuations subside,
causing a smooth change from the high-temperature phase
in which chiral symmetry is approximately restored towards the normal phase.
Of practical interest is the fact that the equilibrium field configurations
can be sampled in a simple manner,
thus providing a convenient means for specifying the initial conditions
in dynamical simulations of the non-equilibrium relaxation of the chiral field.
The corresponding correlation function is briefly considered
and used to calculate the spectral strength of radiated pions.
Finally, by propagating samples of initial configurations
by the exact equation of motion,
it has been ascertained that the treatment is sufficiently accurate
to be of practical utility.

\vfill
\noindent
$^*$This work was supported by the Director,
Office of Energy Research,
Office of High Energy and Nuclear Physics,
Nuclear Physics Division of the U.S. Department of Energy
under Contract No.\ DE-AC03-76SF00098,
and by the National Institute for Nuclear Theory
at the University of Washington in Seattle.\\
\noindent
}
\end{titlepage}

\section{Introduction}
\label{introduction}

The possibility of producing and observing disoriented chiral condensates
in high-energy collisions of hadrons and nuclei has stimulated considerable
interest over the past few years
(for a recent review, see ref.\ \cite{Rajagopal:QGP2,Gavin:QM95}).
The basic premise is that the collision generates an extended domain of space
within which  chiral symmetry is approximately restored.
If this happens,
macroscopic pion fields may be generated as a consequence of
the subsequent non-equilibrium relaxation towards the normal state.
Such isospin-aligned domains may manifest themselves in anomalous pion emission
\cite{Anselm,Blaizot:PRD46,Rajagopal:NPB399}
of the type seen in the Centauro cosmic ray events \cite{Centauro}.
The experimental exploration of this phenomenon is of fundamental interest
because it has a direct bearing
on the mechanism of spontaneous chiral symmetry breaking.
Efforts are well underway to search for
the associated pion multiplicity fluctuations
in proton-proton collisions \cite{Bjorken}.

In order to assess the prospects for such a phenomenon to actually occur
and be detectable above the background of other pion production processes,
it is necessary to perform extensive dynamical calculations.
This is a daunting task
because the chiral degrees of freedom should be properly embedded
in the complicated environment generated in a high-energy collision
which evolves from being primarily partonic at the early stage
to entirely hadronic in the course of the chiral relaxation process.
Fortunately,
the study of suitable idealized scenarios can yield valuable insights
regarding the prospects for observing the effect.

Most dynamical studies have been carried out
within the framework of the linear $\sigma$ model,
treated in the mean-field approximation
\cite{Rajagopal:NPB404,GGP93,GM,Kluger,Blaizot:PRD50,%
Huang:PRD49,Asakawa,Boyanovsky,Csernai,MM}.
The physical problem divides naturally into three parts:
1) specification of the initial state
in which chiral symmetry is approximately restored;
2) the dynamical evolution during which the chiral field cools
and evolves towards the neighborhood of the physical vacuum; 
and 3) the analysis of the asymptotic state in terms of physical observables.
The present work is particularly relevant for the first issue,
as it provides a well-based framework
for characterizing the chiral field configurations
that are employed as initial states in the numerical simulations.
In particular,
a convenient method is developed for sampling
the statistical equilibrium ensemble of chiral configurations
and is expected to be of practical utility.

After a brief reminder of the relevant features of
the linear $\sigma$ model (sect.\ \ref{model}),
we describe how thermal equilibrium can be treated approximately
by means of a standard linearization procedure (sect.\ \ref{equil}).
We then discuss and illustrate the statistical distribution
of the average field strength, the order parameter (sect.\ \ref{order}),
and subsequently turn to the properties of
the quasi-particle degrees of freedom
associated with the spatial field fluctuations (sect.\ \ref{qp}).
An impression of the validity of the treatment is then gained
by evolving samples of field configurations by the exact equation of motion
(sect.\ \ref{dynamics}).
Finally, a concluding discussion is given (sect.\ \ref{conclusion}).

\newpage
\section{The linear ${\sigma}$ model}
\label{model}

To set the framework for the subsequent developments,
we start by briefly recalling the most relevant features
of the formal framework.
The present study is carried out within the mean-field approximation
where the quantum field operators are replaced by their expectation values,
thereby bringing the treatment to the level of classical field theory.
This simplified treatment is expected to suffice for exploratory calculations.
Naturally,
the mean-field approximation is only a first step
towards a more complete description.

The basic object of study is then the chiral field
\beq
\Phi(\r,t)=\sigma(\r,t)+ i\bold{\tau}\cdot\bold{\pi}(\r,t)\ ,
\eeq
where the three elements of the vector $\bold{\tau}$ are the Pauli matrices.
Here the scalar field $\sigma(\r,t)$
and the vector field $\bold{\pi}(\r,t)$
are both real and can conveniently be combined into
the $O(4)$ vector $\pphi=(\sigma,\bold{\pi})$.

In the linear $\sigma$ model \cite{sigma-model},
one introduces a simple local effective interaction energy density,
\beq\label{V}
V(\pphi)={1\over\hbar^3c^3}\left[{\lambda\over4}(\phi^2-v^2)^2-H\sigma\right]\ ,
\eeq
where $\phi$ denotes the magnitude of the $O(4)$ vector $\pphi$,
\beq
\phi^2\equiv\pphi\cdot\pphi = \sum_{j=0}^3 \phi_j(\r,t)^2
=\sigma^2 +\bold{\pi}\cdot\bold{\pi}\ ,
\eeq
with $j=0,1,2,3$ referring to the four chiral directions.
The interaction $V$ contains three parameters:
$\lambda$, the strength of the symmetric term;
$v$, the location of its minimum;
and $H$, the strength of the symmetry-breaking term.
As is commonly done,
these parameters are fixed by specifying
the pion decay constant, $f_\pi=92\ \MeV$, 
and the meson masses, $m_\pi=138\ \MeV/c^2$ and $m_\sigma=600\ \MeV/c^2$,
leading to
\beqar \label{lambda}
\lambda &=& {m_\sigma^2c^2-m_\pi^2c^2 \over 2f_\pi^2}=20.14\ ,\\ \label{v}
v &=&
\left[{m_\sigma^2-3m_\pi^2\over m_\sigma^2-m_\pi^2}f_\pi^2\right]^{1\over2}
=86.71\ \MeV\ , \\ \label{H}
H &=& m_\pi^2c^2 f_\pi= (120.55\ \MeV)^3\ .
\eeqar
The precise values of the model parameters are immaterial,
both in view of the simple nature of the model
and, in particular, within the context of our present idealized study.
The value for the hypothetical $\sigma$ mass is the most commonly adopted one;
for $m_\pi$ we have used the weighted average of the three observed pion masses,
and for the pion decay constant we have simply taken two thirds of the pion mass
which gives a simple value within the range of experimental data
(the current values being $f_{\pi^\pm}=92.4\pm0.26\ \MeV$ and
$f_{\pi^0}=84\pm3\ \MeV$ \cite{DATA}).

The Lagrangian density is given by
\beq\label{L}
{\cal L}(\r,t)={1\over\hbar^3c^3}\
\left[{1\over2}(\hbar\del_t\pphi)^2
-{1\over2}(\hbar c \nabla\pphi)^2
-{\lambda\over4}(\phi^2-v^2)^2+H\sigma\right]\ .
\eeq
The corresponding energy density is then
\beq\label{Edens}
{\cal H}(\r,t)={1\over\hbar^3c^3}\
\left[{1\over2}\psi^2
+{1\over2}(\hbar c \nabla\pphi)^2
+{\lambda\over4}(\phi^2-v^2)^2-H\sigma\right]\ ,
\eeq
where the time derivative  $\ppsi\equiv\hbar\del_t\pphi$
is the canonical conjugate
of the field strength.  

The action generated by a given time evolution is given by
${\cal S}=\int d\r dt{\cal L}(\r,t)$.
By demanding that $\cal S$ be stationary with respect to arbitrary variations
of both the field strength $\pphi(\r,t)$ and its first derivatives,
one obtains the associated equation of motion,
\beq\label{EoM}
\Box\pphi+\lambda(\phi^2-v^2)\pphi=H\hat{\bold{\sigma}}\ ,
\eeq
where $\Box=\hbar^2\del_t^2-\hbar^2c^2\triangle$
is the d'Alambert differential operator,
and $\hat{\bold{\sigma}}$ denotes the unit vector in the $\sigma$ direction. 
Since the equation of motion is of second order in time,
the evolution of the system is fully determined
once the initial values of the field strength $\pphi(\r)$
and the associated time derivative $\ppsi(\r)$ have been specified.

\section{Approximate treatment}
\label{equil}

We now discuss how the chiral system can be treated in a convenient
approximate manner.
For this purpose,
we confine the system within a rectangular box and impose periodic boundary
conditions.

\subsection{Linearization}

Our starting point is the equation of motion (\ref{EoM}) for the field strength.
It is natural to decompose the field,
\beq
\pphi(\r,t)=\ul{\pphi}(t) +\delta\pphi(\r,t)\ ,
\eeq
where the first term is the spatial average,
$\ul{\pphi}\equiv<\pphi>$, so that $<\delta\pphi>=\bold{0}$.
The average part of the field, $\ul{\pphi}$,
is often referred to as the order parameter,
whereas the spatial fluctuations, $\delta\pphi(\r)$,
may be considered as quasi-particle degrees of freedom
representing elementary excitations 
relative to the constant field configuration \cite{GM}.
The $O(4)$ direction of the order parameter plays a special role
and it it convenient to employ a correspondingly aligned reference system,
\ie\ a system in which $\ul{\pphi}=(\phi_0,\bold{0})$
and $\delta\pphi=(\dphip,\delta\pphit)$.
The thermal average $\prec\delta\pphi\ \delta\pphi\succ$ is then
a diagonal $4\times4$ tensor,
as can be demonstrated by elementary means.

Taking the spatial average of the full equation of motion (\ref{EoM})
leads to an equation for time evolution of the order parameter $\ul{\pphi}$
\cite{Baym:PRD15},
\beq\label{phibar}
\hbar^2\del_t^2\ul{\pphi} +\lambda
[\phi_0^2 +\prec\delta\phi^2\succ +2\prec\dphip^2\succ-v^2]\ul{\pphi}
=H\hat{\bold{\sigma}}\ ,
\eeq
In deriving this equation,
we have replaced the spatial averages $<\cdot>$ by thermal averages
$\prec\cdot\succ$,
as may be justified for a system of dimensions larger
than the correlation length
(the fluctuations in different regions of the system are then independent
and can therefore be assumed to reflect the thermal distribution
from which the particular field was sampled). 
We shall use $\prec\dphit^2\succ$ to denote the diagonal elements
of the isotropic $3\times3$ tensor $<\delta\pphit\ \delta\pphit>$,
so that $\prec\delta\phi^2\succ=\prec\dphip^2\succ+3\prec\dphit^2\succ$
is the total fluctuation.\footnote{We call attention
to the fact that in the present notation $\prec\dphit^2\succ$ denotes the
variance of the field fluctuations in one of the transverse directions,
whereas in refs.\ \cite{GM,Asakawa} it represents three times that amount,
namely the sum of the variances in all three perpendicular directions.}
Furthermore,
$<\delta\phi^2\ \delta\pphi>$ vanishes by symmetry.

By subtracting the equation of motion for the order parameter,
eq.\ (\ref{phibar}),
from the full equation of motion (\ref{EoM}),
it is possible to obtain approximate equations for the field fluctuations,
\beqar\label{EOMp}
\left[ \Box + \mup^2 c^4 \right] \dphip &=& 0\ ,\\ 
\label{EOMt}
\left[ \Box + \mut^2 c^4\right] \bold{\dphit} &=& \bold{0}\ ,
\eeqar
where the effective masses $\mup$ and $\mut$
are determined by the auxiliary gap equations,
\beqar\label{mup}
\mup^2 c^4 &\equiv& \lambda
\left( 3\phi_0^2 +\prec\dphi^2\succ +2\prec\dphip^2\succ -v^2 \right)\ ,\\
\label{mut}
\mut^2 c^4 &\equiv& \lambda
\left( \phi_0^2 +\prec\dphi^2\succ +2\prec\dphit^2\succ -v^2 \right)\ .
\eeqar

In arriving at this result,
we have replaced products of two individual fluctuations
by their respective thermal ensemble average values,
$\delta\phi_j\delta\phi_{j'} \leadsto\ \prec\delta\phi_j\delta\phi_{j'}\succ$,
and, furthermore, products containing three fluctuation factors
have been contracted in the usual manner \cite{Boyanovsky},
\beq
\delta\phi_j\delta\phi_{j'}\delta\phi_{j''}\ \leadsto\
\prec\delta\phi_j\delta\phi_{j'}\succ\delta\phi_{j''}\ +\
\prec\delta\phi_{j'}\delta\phi_{j''}\succ\delta\phi_j\ +\
\prec\delta\phi_{j''}\delta\phi_j\succ\delta\phi_{j'}\ .
\eeq

Equations (\ref{EOMp}) and (\ref{EOMt})
describe the fluctuations as independent
quasi-particles having the respective effective masses $\mup$ and $\mut$,
which in turn are given in terms of the field fluctuations,
eqs.\ (\ref{mup}) and (\ref{mut}).
This self-consistent relationship can be stated explicitly by invoking
the expression for the associated thermal equilibrium fluctuations,
\beq\label{fluct}
\prec\dphip^2\succ\ =\ {\hbar^3c^3\over\Omega}
{\sum_\k}' {1\over\eps_k}{1\over\rme^{\eps_k/T}-1}\
\asymp\ {T\over2\pi^2}\ \mup c^2
\sum_{n>0} {1\over n} K_1(n{\mup c^2\over T})\ ,
\eeq
and similarly for $\prec\dphit^2\succ$.
The system is enclosed in a rectangular box with
the volume $\Omega=L_xL_yL_z$
and periodic boundary conditions are imposed.
The wave vectors $\k$ are then quantized
($k_xL_x = 2\pi K_x$ with $K_x=0,\pm1,\dots$, \etc),
so that the individual quasi-particle modes are enumerated by $(K_x,K_y,K_z)$.
The corresponding quasi-particle energy is determined by
the dispersion relation $\eps_k^2=\hbar^2c^2k^2+\mu^2c^4$,
where the effective masses $\mup$ and $\mut$
lead to the energies $\eps_k^\parallel$ and $\eps_k^\perp$, respectively.
Finally, the prime on the summation sign indicates that
the mode having $k=0$ should be omitted
since it represents the order parameter rather than a spatial fluctuation.

In eq.\ (\ref{fluct}),
the symbol $\asymp$ indicates the thermodynamic limit $L\to\infty$
in which the quasi-particle modes form a continuum.
The fluctuations can then be expressed analytically
in terms of the Modified Bessel Function $K_1$, as indicated.
Generally,
the fluctuations decrease as the effective mass increases,
for a fixed temperature $T$,
and $\prec\dphi_j^2\succ=T^2/12$ when the effective mass $\mu_j$ vanishes.
The global approximation
$\prec\delta\phi_j^2\succ\approx (T^2/12)(1+\mu c^2/2\pi T)\exp(-\mu c^2/T)$
is good to better than 2\%.

The treatment above has been carried out under the simplifying assumption
of thermal equilibrium which suffices for our present purposes.
However, it is interesting to note that the results can be readily
generalized to non-equilibrium scenarios by simply replacing the thermal
averages $\prec\cdot\succ$ by the corresponding spatial averages $<\cdot>$.
In this manner, the system of equations would be closed
and a conceptually simple dynamical description emerges.
However,
since the direct interaction between the quasi-particles is then ignored,
any relaxation can occur only via the mean field,
and the accuracy of such an approach should therefore be carefully assessed.

\subsubsection{Symmetries}

It is important to note that the equation of motion (\ref{phibar})
for the spatial average of the field strength,
the order parameter $\ul{\pphi}$,
has four components,
one for each of the chiral directions.

In the special case when the symmetry-breaking term vanishes, $H=0$,
the interaction term in (\ref{phibar}) depends only on
the magnitude of the order parameter, $\phi_0$.
The evolution of $\ul{\pphi}$ is then equivalent to that of
a particle in a central field in four dimensions and,
accordingly,
the dynamics separates into one radial and three angular modes.
The rotational invariance implies that the angular motion is massless,
a manifestation of the Goldstone Theorem (see, for example, ref.\ \cite{Zuber}).
The conserved currents associated with this $O(4)$ invariance are
the Noether iso-vector current 
$\bold{V}_\mu=\bold{\pi}\times\del_\mu\bold{\pi}$
and iso-axial-vector current 
$\bold{A}_\mu=\bold{\pi}\del_\mu\sigma-\sigma\del_\mu\bold{\pi}$
(related to rotations around the $\sigma$ axis
and rotations of the $\sigma$ axis,
respectively).
In the general case, when $H$ does not vanish,
the model remains invariant with respect to rotations around the $\sigma$ axis
and the associated $O(3)$ symmetry ensures the conservation of isospin.

We also note in this connection that the quasi-particles modes
all have a finite effective mass (except right at the critical point).
This should pose no puzzle since they have always finite momenta,
whereas the above modes carry no momentum.

\subsection{Effective masses}
\label{masses}

Utilizing the result (\ref{fluct}),
the coupled equations (\ref{mup}--\ref{mut}) for the effective masses
can be solved for specified values of the temperature $T$
and the magnitude $\phi_0$,
provided that these parameters are sufficiently large,
and we then have $\mup \geq \mut \geq 0$.
It should be noted that the effective masses
are independent of the parameter $H$,
since the symmetry-breaking term $H\sigma$ is linear.\footnote{This simplifying
feature holds only when $\lambda$, $v$, and $H$ are considered as
the primary model parameters.
When $H$ vanishes it is often customary to readjust the other two parameters,
$\lambda=m_\sigma^2c^2/2f_\pi^2=21.27$ and $v=f_\pi=92\ \MeV$,
and then the effective masses change correspondingly.}

In fig.\ \ref{f:mu} we show the resulting effective masses $\mup$ and $\mut$
as functions of $\phi_0$,
for temperatures $T$ up to well above critical.
At any temperature,
there is always a physical solution to the coupled equations
(\ref{mup}-\ref{mut}) for the effective masses when $\phi_0\geq v$.
This is easy to see from eq.\ (\ref{mut}):
At $T=0$, when the fluctuations vanish,
we have $\mut^2=\lambda(\phi_0^2-v^2)$
and so $\mut^2$ vanishes at $\phi_0=v$ and is positive for larger $\phi_0$;
an increase of $T$ will always increase the fluctuations,
and hence the mass.
Moreover, we always have $\mup\geq\mut$.

Since the field fluctuations and the magnitude of the order parameter
contribute to the effective masses in qualitatively similar ways,
an increase of the temperature (and thus the fluctuations)
will permit a further decrease of $\phi_0$, 
so that the point at which $\mut$ vanishes
is moved to ever smaller values of $\phi_0$.
The behavior of $\mup$,
considered as a function of $\phi_0$,
is nearly independent of temperature (as long as $T$ is subcritical),
except that each curve terminates
at the point where the corresponding $\mut$ vanishes.
This limiting curve is indicated by the dotted curve on the interval $(0,v)$
and it is elementary to calculate,
\beqar
\phi_0^2 &=& v^2-{5\over12}T^2-\prec\dphip^2\succ\
\approx\ v^2-{T^2\over12}(5+\rme^{-\mup c^2/T})\ ,\\
\mup^2c^4 &=& \lambda(2v^2-T^2)\ .
\eeqar

When the temperature reaches the critical value, $T=T_c$,
there is a solution to eqs.\ (\ref{mup}-\ref{mut}) for all values of $\phi_0$,
and both effective masses vanish for $\phi_0=0$.
Then $\prec \delta\phi_j^2 \succ=T_c^2/12$ and so $T_c^2=2v^2$,
\ie\ $T_c=122.63\ \MeV$ with the adopted parameter values.
The degeneracy in the masses, $\mup=\mut$, is a general consequence of the 
$O(4)$ rotational symmetry that emerges for $\phi_0=0$
and it therefore remains as $T$ is further increased,
with the common mass value $\mu_0$ increasing steadily.
Since the effective mass at $\phi_0=0$ is given by 
$\mu_0^2c^4=\lambda(6\prec \delta\phi_j^2 \succ -v^2)$,
it becomes proportional to $T$ at high temperatures,
$\mu_0 c^2\approx1.59\ T$ for $T\gg v$.

The results displayed in fig.\ \ref{f:mu} have been calculated
in the thermodynamic limit, $L\to\infty$,
in which the quasi-particle modes form a continuum.
However,
the systems of interest in connection with high-energy nuclear collisions
have finite volumes.
The quasi-particle energies are then discrete and, as a result,
the effective masses are modified.
This effect is largest when the effective mass is small,
\ie\ near the critical conditions,
because the absence of the constant mode in (\ref{fluct})
is then most significant,
and it generally leads to smaller masses.
Figure \ref{f:muL} shows the temperature dependence of the effective masses
for either $\phi_0=0$
or for $\phi_0=f_\pi$ (the value associated with the physical vacuum),
as obtained either in the continuum limit
or for a cubic box with a side length $L$ of either 8 fm or 5 fm.

For $\phi_0\approx0$ and temperatures near critical,
the effective masses are significantly reduced.
As a consequence,
the associated critical temperature is increased
to 154 MeV for $L=8\ \fm$ and to 172 MeV for $L=5\ \fm$.
However,
for temperatures several tens of MeV above these thresholds,
the effect is relatively small.
The effect also disappears quickly
as the order parameter is moved away from zero
and by the time it reaches its vacuum value, $\phi_{\rm vac}=f_\pi=92\ \MeV$,
there is hardly any effect for even the smallest box ($L=5\ \fm$).

The effect of the finite size on the fluctuations in the field strength
is illustrated in fig.\ \ref{f:phi2}.
The upper curves show the typical magnitude of the thermal fluctuations,
$(\prec\dphi^2\succ)^{1/2}$,
as functions of $\phi_0$ over a range of temperatures,
for the same three box sizes as in fig.\ \ref{f:muL},
while the lower curves show the continuum result
for the contribution arising from fluctuations directed along
the order parameter $\ul{\pphi}$, $\prec\dphip^2\succ^{1/2}$.
The fluctuations are remarkably independent of the volume,
except in the critical region $T\approx T_c$ and $\phi_0\approx0$,
as just discussed.
For $\phi_0=0$ the $O(4)$ symmetry implies that the fluctuations
are the same in all directions
so that each of the lower curves starts out at
half the value of the corresponding upper curve.
For larger values of $\phi_0$
the relative contribution from fluctuations along $\ul{\pphi}$
decreases progressively,
since $\mup$ grows much larger than $\mut$ (see fig.\ \ref{f:mu}).

\subsection{Energy}
\label{energy}

It is instructive to examine the total energy of the system.
It is obtained by integrating the energy density (\ref{Edens})
over the volume of the box,
\beqar\label{Etot}
{\cal E}(t) &=& \int d\r\ {\cal H}(\r,t)\
=\ \Omega<{1\over2}\hbar^2\psi^2 +{1\over2}\hbar^2c^2(\nabla\pphi)^2 +V>\\
\label{Eterms}
&=&\ \Omega(E_0+E_{\rm qp}+\delta V)\ .
\eeqar
The total energy is conserved in time
when the field is evolved with the equation of motion (\ref{EoM}).
The energy is composed of the three basic contributions
exhibited in the second expression:
one arising from the time dependence of the field strength,
another associated with the spatial variation of the field,
and a third resulting from the interaction.
In the last expression,
these contributions have been reorganized into three different terms
which we will now discuss in turn.

The first term is the bare energy density,
\ie\ the energy density that would arise if the field had no spatial variation,
$\pphi=\ul{\pphi}$ and $\ppsi=\ul{\ppsi}$,
\beq\label{E0}
E_0={1\over\hbar^3c^3}
\left[{1\over2}\psi_0^2 
+{\lambda\over4}(\phi_0^2-v^2)^2-H\phi_0\cos\chi_0\right]
=K_0+V_0\ .
\eeq
We have here introduced $\chi_0$ to denote the angle of disorientation,
\ie\ the angle between the four-dimensional order parameter $\ul{\pphi}$
and the $\sigma$ direction.
The spatial average of $\sigma$ is then
$\sigma_0=\ul{\pphi}\cdot\hat{\bold{\sigma}}=\phi_0\cos\chi_0$,
and the corresponding mean magnitude of the pion field is
$\pi_0=\phi_0\sin\chi_0$.
In the last expression $K_0$ denotes the kinetic part of $E_0$
and $V_0$ is the bare potential energy density.

The second term in (\ref{Eterms}) denotes the energy density
associated with the quasi-particle degrees of freedom,
in the approximation that these form a gas of independent particles,
\beqar\label{Eqp}
E_{\rm qp}&=&{1\over\hbar^3c^3}\sum_{j=0}^4{1\over2}
<(\hbar\del_t\dphi_j)^2 +(\hbar c\nabla\dphi_j)^2 + \mu_j^2c^4\dphi_j^2>\\
&=&{1\over\hbar^3c^3}\sum_{j=0}^4 {1\over2}{\sum_\k}'
(|\psi_\k^{(j)}|^2 +(\hbar^2k^2c^2+\mu_j^2c^4)|\phi_\k^{(j)}|^2)\ .
\eeqar
We have here invoked the general Fourier expansions
of $\pphi(\r)$ and $\ppsi(\r)$,
\beqar
\label{phik}    \pphi(\r,t)&=&\sum_\k \pphi_\k(t)\ \rme^{i\k\cdot\r}\ ,\\
\label{psik}    \ppsi(\r,t)&=&\sum_\k \ppsi_\k(t)\ \rme^{i\k\cdot\r}\ .
\eeqar
The expansion coefficients are generally complex and are
given simply in terms of spatial averages,
$\pphi_\k=<\pphi(\r)\exp(-i\k\cdot\r)>$ and
$\ppsi_\k=<\ppsi(\r)\exp(-i\k\cdot\r)>$.
We note that the components with $\k=\bold{0}$ represent the order parameter,
$\pphi_{\bold{0}}=\ul{\pphi}$ and $\ppsi_{\bold{0}}=\ul{\ppsi}$.
Furthermore, the fact that the chiral field is real imposes the relations
$\pphi_\k^*=\pphi_{-\k}$ and $\ppsi_\k^*=\ppsi_{-\k}$.

The third term in (\ref{Eterms}) corrects for the fact that the quasi-particles
in fact interact via the fourth-order term in $V$,
\beq\label{dV}
\delta V\ =\ {\lambda\over4}
[<\dphi^4>\ -\ 2<\dphi^2>^2\ -\ 4<\sum_{j=0}^3\dphi_j^2>^2]\
\approx\ -{\lambda\over4}<\dphi^4>\ .
\eeq
The last result can be obtained by using the Gaussian approximation
to express $<\dphi^4>$ in terms of squares of quadratic terms.

The energy of the system can be calculated as a function of the
order parameter,
assuming that the quasi-particle degrees of freedom are in thermal equilibrium.
The resulting energy density is
\beq\label{Ebar}
\ul{E}(\ul{\ppsi},\ul{\pphi})=
 E_0(\psi_0,\phi_0,\chi_0)\ +\ \prec E_{\rm qp}\succ(\phi_0)\
+\ \prec\delta V(\phi_0)\succ
=K_0+V_T(\phi_0,\chi_0)\ .
\eeq
Here the thermal equilibrium value of the quasi-particle energy density
is given by
\beq\label{EqpT}
\prec E_{\rm qp}^{(j)}\succ\ =\
{1\over\Omega} {\sum_\k}' {\eps_k\over{\rme^{\eps_k/T}-1}}\
\asymp\ {(\mu c^2)^4\over\hbar^3c^3}{1\over2\pi^2}
\sum_{n>0} \left[3{\tau^2\over n^2} K_2({n\over\tau})
+{\tau\over n} K_1({n\over\tau})\right]
\eeq
for each of the four chiral directions $j$, with $\tau\equiv T/\mu_j c^2$.
It can be expressed analytically in the continuum limit, as indicated.
Furthermore,
the thermal equilibrium value of the correction term is quite well
represented by its Gaussian approximation,
\beq
\delta V\ \approx\ -{\lambda\over4}
[\prec \dphi^2\succ^2\
+\ 2\prec \dphi_\parallel^2\succ^2\ +\ 6\prec \dphi_\perp^2\succ^2]\ .
\eeq

In the last expression in (\ref{Ebar})
the energy has been split into the kinetic energy
associated with the time dependence of the order parameter,
$K_0=\psi_0^2/2\hbar^3c^3$,
and the remainder which can be regarded as an effective potential,
\beq
V_T(\phi_0,\chi_0)\ =\ V_0\ +\ \prec E_{\rm qp}\succ\ +\ \prec\delta V\succ\ .
\eeq
This effective potential has been plotted in fig.\ \ref{f:V}
for the particular situation when the order parameter $\ul{\pphi}$
is directed along the $\sigma$ axis,
for a range of temperatures $T$.
The directional dependence of $V_T$
is simply given by the term $H\phi_0\cos\chi_0/\hbar^3c^3$.
At zero temperature
the effective potential can only be displayed for order parameters 
whose magnitude $\phi_0$ exceed $v$,
since only then can the effective masses be calculated,
as explained in sect.\ \ref{masses}:
For smaller values of $\phi_0$ some of the low-energy quasi-particle modes
are unstable and so have no well-defined thermal equilibrium.
As the temperature is increased,
the domain of stability extends ever lower towards $\phi_0=0$.
The limiting curve connecting the end points of all the subcritical curves
is determined by the condition that $\mut$ vanish,
implying $\mup^2=\lambda(2v^2-T^2)$.
The result is shown on the figure as the dotted curve
between $\phi_0=0$ and $\phi_0=v$.

For $H=0$ (fig.\ \ref{f:V}$a$) the effective potential has $O(4)$ symmetry
and so the minima form the surface of a 4-sphere;
its intersections with the $\sigma$ axis are indicated by the two arrows.
The addition of the linear symmetry-breaking term (fig.\ \ref{f:V}$b$) 
tilts the potential towards the $\sigma$ direction and,
consequently, there is only one minimum for any given temperature
and it is located on the $\sigma$ axis.

For all values of $H$,
the minima of the effective potential move gradually towards larger
values of the order parameter.
In this respect, the results are similar to the behavior of the
energy of nuclear matter as a function of the density $\rho$
(see, for example, ref.\ \cite{Emil-1}).
However,
while the energetics alone thus favors ever larger values
of either $\phi_0$ or $\rho$ as the temperature is raised,
the proper incorporation of the phase space
(by means of the appropriate entropy)
turns the picture around, as we shall now see.

\section{The order parameter}
\label{order}

We can now consider the distribution
of the chiral field in statistical equilibrium at a given temperature $T$.
Since the equation of motion (\ref{EoM}) is of second order in time,
we need to specify both the field strength $\pphi(\r)$
and its time derivative $\ppsi(\r)\equiv \hbar\del_t\pphi(\r)$
in order to fully characterize a state.
As we have already exploited above,
the quasi-particle degrees of freedom can be regarded as a Bose-Einstein gas 
having thermal distributions determined by their respective effective masses,
in the appropriate domain of $\phi_0$.
Their behavior can then be calculated 
once the order parameter has been specified.

\subsection{The partition function}

It is natural to start the discussion of statistical properties by considering
the partition function for the chiral field.
It is obtained by summing the canonical weights for all the possible states,
\beqar\label{ZT}
{\cal Z}_T&\equiv&\int{\cal D}[\ppsi,\pphi]\
\rme^{-{\Omega\over T}{\cal E}[\ppsi,\pphi]}\
=\prod_\k \left[ \int d^4\ppsi_\k\ d^4\pphi_\k \right]
\rme^{-{\Omega\over T}(E_0 +\delta V +E_{\rm qp})}\\ \label{WT}
&\approx&\int d^4\ul{\ppsi}\ d^4\ul{\pphi}\
\rme^{-{\Omega\over T}(E_0+\prec\delta V\succ)}{\cal Z}_{\rm qp}
=\ \int d^4\ul{\ppsi}\ d^4\ul{\pphi}\ W_T(\ul{\pphi},\ul{\ppsi})\ ,
\eeqar
where $\cal E$ is the total energy of the system
given in eqs.\ (\ref{Etot}-\ref{Eterms}).
The functional integral over all states $(\ppsi(\r),\pphi(\r))$
has first been expressed as a multiple integral
over the four-dimensional Fourier amplitudes $\ppsi_\k$ and $\pphi_\k$ 
(see eqs.\ (\ref{phik}-\ref{psik})).
If the integration over the field fluctuations
(\ie\ those amplitudes having $\k\neq\bold{0}$)
is performed, the remaining integrand is the statistical weight $W_T$
expressing the relative probability for finding the system
with a specified value of the order parameter.
In order to derive an expression for $W_T$,
we have replaced the correction term $\delta V$ (eq.\ (\ref{dV}))
by its approximate thermal average $\prec\delta V\succ$,
for each value $(\ul{\ppsi},\ul{\pphi})$ of the order parameter
(it depends only on $\phi_0$).
The integral over the quasi-particle degrees of freedom can then be performed
and yields the quasi-particle partition function ${\cal Z}_{\rm qp}$.

This quantity factorizes into contributions
from each of the four chiral directions,
${\cal Z}_{\rm qp}=\prod_j {\cal Z}_{\rm qp}^{(j)}$,
where
\beq
{\cal Z}_{\rm qp}^{(j)}\ \equiv\
{\prod_\k}'\left[ \int d\psi^{(j)}_\k\ d\phi^{(j)}_\k\right]
\rme^{-E_{\rm qp}^{(j)}/T}\ \sim\
{\prod_\k}'\left[\int_0^\infty \eps_k  dC_\k^2\int_0^{2\pi} d\eta_\k\right]
\rme^{-E_{\rm qp}^{(j)}/T}\ .
\eeq
Since the complex Fourier amplitudes
$\psi^{(j)}_\k$ and $\phi^{(j)}_\k$
are subject to symmetry relations under $\k\to-\k$,
as already noted above,
is it more convenient to use the representation
in terms of trigonometric functions, eqs.\ (\ref{dphi}) and (\ref{dpsi}).
The different wave numbers are then fully decoupled,
and for each $\k$ the integration variables are
the (positive) amplitude $C_\k$ and the associated phase $\eta_\k$,
as indicated in the last expression above.
The phases are unimportant here since they each merely contribute
a factor of $2\pi$.
Moreover, the combination $n_\k\equiv\eps_k C_\k^2/\hbar^3c^3$
can be interpreted as the average number of quanta in the mode $\k$
(see eq.\ (\ref{nk})).
Since we want to take account of the quantal nature
of the quasi-particle degrees of freedom (see \ref{occup}),
we replace the continuous integral over $n_\k$
by a discrete sum over the possible integer values of the occupation number,
thereby obtaining the standard expression
for an ideal Bose-Einstein gas,
\beq
{\cal Z}_{\rm qp}^{(j)}\ \sim\
{\prod_\k}'\left[\sum_{n_\k=0}^\infty\rme^{-n_\k \eps_k/T}\right]\ =\
{\prod_\k}'(1-\rme^{-\eps_k/T})^{-1}\ =\ {\prod_\k}'\fbar_k^{(j)}\ .
\eeq
We have used the fact that the quasi-particle energy is additive,
$E_{\rm qp}^{(j)}=\sum_\k n_\k \eps_k$.
Furthermore,
$\fbar_k\equiv1+f_k$ where $f_k=(\rme^{\eps_k/T}-1)^{-1}=\prec n_\k\succ$
is the mean number of quanta (in the chiral direction $j$)
having a wave number of magnitude $k$,
and the corresponding energy $\eps_k$ is determined by
$\eps_k^2=\hbar^2c^2k^2+\mu_j^2c^4$.

It follows that the free energy density of the quasi-particles is
$F_{\rm qp}=\sum_j F_{\rm qp}^{(j)}$ with
\beq
F_{\rm qp}^{(j)}\equiv -{T\over \Omega}\ln{\cal Z}_{\rm qp}^{(j)}
={\sum_\k}'\left[\eps_k f_k -T(\fbar_k\ln\fbar_k -f_k\ln f_k)\right]
=\prec E_{\rm qp}^{(j)}\succ -T S_T^{(j)}\ .
\eeq
In the resulting expression,
$\prec E_{\rm qp}^{(j)}\succ$ is the average thermal energy density
of the quasi-particle modes associated with the chiral direction $j$,
eq.\ (\ref{EqpT}), and $S_T^{(j)}$ is their entropy density.
The total entropy density at the temperature $T$ is then
$S_T=S_T^\parallel+3S_T^\perp$,
where $S_T^\parallel(\phi_0)$ represents the entropy density
of the field fluctuations along the direction of $\ul{\pphi}$ \cite{KuboI},
\beqar\label{S}
S_T^\parallel&=&{1\over\Omega}{\sum_\k}'
\left[\fbar_k^\parallel \ln\fbar_k^\parallel
-f_k^\parallel\ln f_k^\parallel \right]\\
&\asymp& {1\over\hbar^3c^3}
\int_{\mu_\parallel c^2}^\infty {d\eps\over2\pi^2}\ \eps\
(\eps^2-\mup^2c^4)^{1\over2}
\left[\fbar(\eps) \ln\fbar(\eps) -f(\eps)\ln f(\eps) \right]\ .
\eeqar
and the contribution to the entropy density
from fluctuations in a perpendicular direction,
$S_T^\perp(\phi_0)$, can be calculated analogously
by replacing $\mup$ with $\mut$.

It is evident from fig.\ \ref{f:V}
that the energetics favors rather large magnitudes of the order parameter,
as would be expected from the form of the interaction potential $V$.
However,
the density of quasi-particle states is higher at small values of $\phi_0$,
since the effective masses are then smaller,
for a given temperature.\footnote{The importance of the quasi-particle entropy
in determining the statistical properties of the model
was already emphasized by Bardeen and Moshe \cite{Moshe:PRD34} a decade ago.}
This interplay between energy and phase-space is quantified
in the statistical weight (see eq.\ (\ref{WT})),
\beqar\label{W}
W_T(\ul{\ppsi},\ul{\pphi}) &\equiv&
\int{\cal D}[\ppsi',\pphi']\
\rme^{-{\cal E}[\ppsi',\pphi']/T}\
\delta^4(\ul{\ppsi}'-\ul{\ppsi})\ \delta^4(\ul{\pphi}'-\ul{\pphi})\\
&\approx& \rme^{-{\Omega\over T}(K_0+V_0 +\prec\delta V\succ +F_{\rm qp})}
=\rme^{-{\Omega\over T}(K_0+F_T)}
\eeqar
We have used the approximate decomposition (\ref{Eterms}) of the total energy
which causes the partition function to factorize and, in the last expression,
we have introduced the free energy density for the order parameter,
$F_T(\ul{\pphi})=V_T(\phi_0,\chi_0)-TS_T(\phi_0)$.

The free energy density is illustrated in fig.\ \ref{f:F}
for a range of temperatures above which the trend is obvious.
Since the entropy density of a free gas at fixed $T$ goes up
when the particle mass is reduced,
the entropy provides a restoring force towards symmetry
(recall that a reduction of the order parameter $\phi_0$
leads to smaller effective masses). 
As a result,
for $H=0$ and temperatures well above $T_c$
the free energy density then has its minimum at $\phi_0=0$.
As the temperature drops,
the symmetric minimum grows ever more shallow.
Near $T\approx175\ \MeV$
a secondary minimum appears at $\phi_0\approx50\ \MeV$,
and it becomes the lowest one at $T\approx171\ \MeV$.
As $T$ decreases further the minimum gently approaches $\phi_0=v$ .
The abrupt change in the location of the free-energy minimum
is characteristic of a first-order phase transition.
In contrast,
when $H$ takes on a realistic finite value (see eq.\ (\ref{H}))
the minimum of $F_T$ is always located on the positive part of the $\sigma$ axis
and it moves smoothly outwards as the system is cooled,
with no drastic change at any temperature.
The resulting behavior is shown by the solid curve in fig.\ \ref{f:Tphi}.

\subsection{Distributions}

The preceding analysis implies that the statistical weight
$W_T(\ul{\ppsi},\ul{\pphi})$,
which is defined on the eight-dimensional phase space of the order parameter,
in fact depends on only three quantities:
the speed $\psi_0=|\ul{\ppsi}|$, the magnitude $\phi_0=|\ul{\pphi}|$,
and the disalignment angle $\chi_0$.
Moreover,
since the entropy depends only on the effective masses,
which in turn are determined solely by $\phi_0$,
the dependence of the free energy on the direction of $\ul{\pphi}$
arises only through the symmetry-breaking $H$ term
in the interaction $V$ (eq.\ (\ref{V})).
Consequently,
the corresponding projected probability distribution factorizes,
\beqar\label{PT} \nonumber
&~&P_T(\psi_0,\phi_0,\chi_0)\
\equiv\ \int d^4\ul{\ppsi}'\int d^4\ul{\pphi}'\ 
\delta(\psi_0'-\psi_0)\ \delta(\phi_0'-\phi_0)\ \delta(\chi_0'-\chi_0)\
W_T(\ul{\ppsi}',\ul{\pphi}')\\
&~& =\ P_\psi(\psi_0)\ P_\phi(\phi_0)\ P_\chi(\phi_0;\chi_0)\
\sim\ \psi_0^3\ \phi_0^3\ \sin^2\chi_0\ W_T(\psi_0,\phi_0,\chi_0)\ .
\eeqar
Here $\psi_0^3$ and $\phi_0^3$ are the Jacobians
associated with the transformation from the
four-dimensional vectors $\ul{\ppsi}$ and $\ul{\pphi}$
to their magnitudes $\psi_0$ and $\phi_0$, respectively,
and $\sin^2\chi_0$ is the Jacobian
arising from the three-dimensional nature of the space
perpendicular to the order parameter $\ul{\pphi}$.
Furthermore,
the three normalized probability densities are given by
\beqar\label{P}
\label{Ppsi}
P_\psi(\psi_0) &=& {\cal N}_\psi\ \psi_0^3\
\exp[-{\Omega\over\hbar^3c^3}{\psi_0^2\over2T}]\ ,
\\ \label{Pphi}
P_\phi(\phi_0) &=& {\cal N}_\phi\ \phi_0^3\
\exp[-\Omega ({1\over T}V_T(\phi_0,\chi_0=0) - S_T(\phi_0))]\ ,
\\ \label{Pchi}
P_\chi(\phi_0;\chi_0) &=& {\cal N}_\chi(\phi_0)\ \sin^2\chi_0\
\exp[-{\Omega \over \hbar^3c^3} {H\over T} \phi_0(1-\cos\chi_0)]\ .
\eeqar
For the normalization constants we have
${\cal N}^{-1}_\psi=2\kappa^2$,
with the convenient abbreviation $\kappa\equiv T\hbar^3c^3/\Omega$,
and ${\cal N}^{-1}_\chi=(\pi/\xi)I_1(\xi)\exp(-\xi)$,
with $\xi\equiv H\phi_0/\kappa$.

The kinetic part, $P_\psi(\psi_0)$,
is the projection of an isotropic four-dimensional normal distribution
having a total variance of $\prec\psi_0^2\succ=4\kappa$.
The average kinetic energy
associated with the time evolution of the order parameter
is then $\prec K \succ=4T$, as would be expected in four dimensions.
The mean speed is $\prec\psi_0\succ =\sqrt{9\pi\kappa/8}$.

Considered as a function of the four-dimensional order parameter $\ul{\pphi}$, 
the probabilty density has a maximum where the free energy has a minimum,
as is evident from eq.\ (\ref{W}).
However,
the jacobian factor $\sim\phi_0^3$ associated with the projection
from $\ul{\pphi}$ to $\phi_0$
increases the most probable value of the magnitude $\phi_0$.
This effect depends on volume $\Omega$,
as is illustrated in fig.\ \ref{f:Tphi}.
The results for a finite box with $L=8\ \fm$ (dashed curve)
are nearly identical to those obtained for $L\to\infty$,
while $L=5\ \fm$ (dots) leads to a more significant increases in $\phi_0$
at the higher temperatures.
The most rapid increase of $\phi_0$ occurs near $T\approx200\ \MeV$
which is well above the critical temperature $T_c$ (arrow)
at which the effective masses drop to zero.

The directional distribution $P_\chi(\phi_0;\chi_0)$,
which depends parametrically on the magnitude $\phi_0$,
grows broader at high temperature where $\phi_0$ decreases.
In the temperature range considered here,
up to several hundred MeV,
$P_\chi(\chi_0)$ is well approximated by the projection of a
three-dimensional normal distribution
with a total variance of $3/\xi$,
and the average magnitude of the disalignment angle $\chi_0$
is then $\prec\chi_0\succ\approx\sqrt{8/\pi\xi}$.

Although the locations of the minima in the free energy
appear qualitatively different when comparing $H=0$ with $H>0$
(figs.\ \ref{f:F}$a$ and \ref{f:F}$b$, respectively),
the corresponding curves for the most probable $\phi_0$
are rather similar when finite volumes of nuclear size are considered,
due to statistical fluctuations.
This is also illustrated in fig.\ \ref{f:Tphi}.
The thin curve is analogous to the solid curve
and shows the dependence of $\phi_0$ in the thermodynamic limit.
(Recall that the vacuum value is $\phi_0=v=86.71\ \MeV$ for $H=0$.)
At $T\approx171\ \MeV$ the value changes abruptly.
The open dots are analoguos to the solid dots
and show the most probable $\phi_0$ for the box having $L=5\ \fm$.
The bias by the jacobian factor $\phi_0^3$ (see eq.\ (\ref{Pphi}))
causes the behavior to become quite similar to that obtained for $H>0$.
In particular,
there is no discontinuity near $T\approx171\ \MeV$
where the minimum in the free energy jumps
between $\phi_0=0$ and $\approx56\ \MeV$.
We finally note that for relatively low temperatures
the Jacobian causes the most probable value of $\phi_0$
to slightly exceed the vacuum value.

\subsubsection{Distribution of the order parameter}

For any finite volume $\Omega$,
the order parameter the exhibits fluctuations around its most probable value.
The full width at half maximum of the $\phi_0$ distribution 
is shown by the horizontal bars in fig.\ \ref{f:Tphi}
for the smallest box ($L=5\ \fm$) where the fluctuations are the largest.
The relative smallness of the fluctuations
indicates that the order parameter is distributed
within a rather limited range of values.
This feature is further illustrated in fig.\ \ref{f:phi}
which depicts the entire distribution, $P_\phi(\phi_0)$,
over a range of temperatures,
for a cubic box with side length $L=8\ \fm$,
which is our standard scenario.
The solid curves in fig.\ \ref{f:phi} have been obtained by
scaling the continuum results to the finite volume.
The dashed curves show the effect of properly quantizing the problem
(\ie\ summing over the discrete modes rather than integrating).
The effect is very small,
because the distributions are peaked well inside
the respective domains of stability.

For small temperatures,
the distribution is narrowly peaked near the vacuum value
$\phi_{\rm vac}$=$f_\pi$ (indicated by the arrow).
As the temperature is increased,
the distribution broadens and gradually begins to move inwards
towards smaller values of $\phi_0$.
The width of the distribution increases from zero at $T=0$,
exhibits a maximum near $T\approx 220\ \MeV$,
and then slowly shrinks as $T\to\infty$.
The somewhat counterintuitive decrease of the fluctuations
at high temperatures 
is due to by the fact that the growing thermal fluctuations
cause the interaction to become progressively more repulsive,
causing the effective potential $V_T$ to become ever more confining.

We recall that for subcritical temperatures, $T<T_c$,
our treatment can only be carried out
above a temperature-dependent minimum value of $\phi_0$.
As it turns it, this principal limitation is of little practical import.
For example, for $T=80\ \MeV$
(the lowest temperature shown in fig.\ \ref{f:phi})
the distance from the centroid (at $\approx91\ \MeV$)
to the boundary (at $\approx69\ \MeV$)
is over ten times the dispersion of the distribution ($\approx2\ \MeV$).
Thus, for any temperature,
the distribution $P_\phi(\phi_0)$ is sufficiently narrow
to make incursions into the respective unstable regime extremely unlikely.

A more global impression of the statistical distribution of the order parameter
$\ul{\pphi}$ can be obtained by considering a contour diagram
of the projected probability density
$P(\phi_0,\chi_0)\equiv P_\phi(\phi_0)\ P_\chi(\phi_0;\chi_0)$.
Such a plot is displayed in fig.\ \ref{f:contour} for a box with $L=8\ \fm$.
The abscissa is the projection of the order parameter onto the $\sigma$ axis,
$\sigma_0\equiv\ul{\pphi}\cdot\hat{\sigma}=\phi_0\cos\chi_0$,
and the ordinate is the magnitude of its perpendicular component,
$\pi_0\equiv\phi_0\sin\chi_0$.
For each temperature,
the dots show the location of the maximum
and the half-maximum contours are traced out.
The solid contours and dots refer to the continuum treatment,
whereas the dashed contours and the open dots
(shown for two temperatures only)
are obtained with a quantized treatment,
which is seen to have little effect.
The profiles in fig.\ \ref{f:phi} are the projections of $P(\phi_0,\chi_0)$
onto the magnitude $\phi_0$.
It is evident from the above results that
the $O(4)$ symmetry is far from restored
at the temperatures expected in the planned high-energy nuclear collisions.
Instead the order parameter is in fact
distributed within a fairly limited domain,
with both its magnitude $\phi_0$ and its angle of disalignment $\chi_0$
exploring only rather narrow ranges.

\subsubsection{Distribution of the effective masses}
\label{mass-dist}

The effective masses are functions of the magnitude of the order parameter, 
$\mup(\phi_0)$ and $\mut(\phi_0)$.
Consequently,
the thermal fluctuations in the order parameter
will cause the effective masses to fluctuate as well.
Since their distributions may be of some interest (see sect.\ \ref{radiation}),
we consider briefly the corresponding probability densities
for the effective masses,
\beq\label{Pmu}
P_\perp(\mu c^2)\equiv
\int d^4 \ul{\pphi}\ P(\ul{\pphi})\ \delta (\mut(\phi_0)c^2-\mu c^2)\ ,
\eeq
and analogously for $P_\parallel(\mu c^2)$,
where $P(\ul{\pphi})=P_\phi(\phi_0) P_\chi(\phi_0;\chi_0)$
is the probability density for the order parameter $\ul{\pphi}$.
In order to gain an impression of these distributions,
we show in fig.\ \ref{f:Tmu} the most probable effective masses
as functions of the temperature.

The transverse effective mass $\mut$ increases steadily
from its free value $m_\pi$ as the temperature is raised,
whereas the parallel effective mass $\mup$ drops by nearly fifty per cent
before finally turning upwards.
For temperatures above $T\approx300\ \MeV$
the two effective masses are practically degenerate 
and gently approach their asymptotic form $\mu c^2\approx1.59\ T$.
We note that the effective masses always exceed the temperature,
so it is never reasonable to ignore the effective masses.
We also note that the transverse mass has practically no strength
near the free pion mass,
until the temperature is below $T_c$ (see fig.\ \ref{f:pions}).

The parallel effective mass $\mup$ exhibits a pronounced minimum
near $T\approx240\ \MeV$.
It is in this region that the order parameter
exhibits its most rapid evolution with $T$ (see fig.\ \ref{f:Tphi})
and the temperature at which the minimum in $\mup$ occurs may be interpreted
as the effective critical temperature of the model \cite{Pisarski}.

\subsection{Sampling of the order parameter}
\label{sample0}

The above analysis provides a convenient basis for sampling the order parameter
in accordance with the statistical weight $W_T(\ul{\pphi},\ul{\ppsi})$
given in eq.\ (\ref{W}),
and we describe briefly how this can be accomplished in a manner that is
both quick (\ie\ the computational effort is small)
and efficient (\ie\ no effort is expended on rejection).

As noted already, the time derivative $\ul{\psi}$
is governed by a four-dimensional normal distribution
which is isotropic and entirely decoupled from the other degrees of freedom
(see eq.\ (\ref{Ppsi})).
Thus it is elementary to sample this quantity.

The most complicated sampling concerns the magnitude $\phi_0$,
due to the intricate structure of its probability distribution,
as discussed above.
However, the numerical effort required is quite modest.
The most efficient method requires a precalculation of the effective masses
as functions of $\phi_0$, for the particular $T$ of interest.
This is quickly done by proceeding as described in sect.\ \ref{masses}.
The $\chi_0$-independent part of the effective potential, $V_T(\phi_0,0)$,
can then be obtained together with the corresponding entropy $S_T(\phi_0)$.
In effect, the probability distribution for $\phi_0$
can be pretabulated, $P_\phi(\phi_0)$ (see eq.\ (\ref{Pphi})),
and it is then a numerically trivial task to sample $\phi_0$.

Once the magnitude $\phi_0$ has been selected,
it is straightforward to sample the disalignment angle $\chi_0$,
using either the exact form (\ref{Pchi})
or its Gaussian approximation.
In order to orient $\ul{\pphi}$ in the $\bold{\pi}$ subspace,
there remains the task of selecting
the remaining $O(3)$ spherical angles $\vartheta_0$ and $\varphi_0$,
upon which the order parameter is given by
$\ul{\pphi}=(\phi_0\cos\chi_0,\
\phi_0\sin\chi_0\sin\vartheta_0\cos\varphi_0,\
\phi_0\sin\chi_0\sin\vartheta_0\sin\varphi_0,\
\phi_0\sin\chi_0\cos\vartheta_0)$.

\section{The quasi-particles}
\label{qp}

We turn now to the discussion of the quasi-particle degrees of freedom
associated with the spatial variations of the chiral field,
$\delta\pphi(\r)$.
Once the magnitude of the order parameter, $\phi_0$,
has been chosen (see sect.\ \ref{sample0}),
the quasi-particle degrees of freedom are fully characterized,
via the effective masses $\mu_\parallel(\phi_0)$ and $\mu_\perp(\phi_0)$,
and it is then possible to sample them them appropriately.

\subsection{Sampling}
\label{sample}

On the basis of the above developments,
it is possible to devise a simple and efficient method
for performing the statistical sampling of the chiral field,
thus putting the initialization of dynamical simulations
on a formally sound basis.

Since the different quasi-particle modes
can be regarded as effectively decoupled,
the sampling is best done by making an expansion into the elementary modes,
\beq\label{dphi}
\dphi_{\parallel}(\r,t)=\left(2\over\Omega\right)^{1\over2}{\sum_\k}'
C_\k^{\parallel}\ \cos(\k\cdot\r-\omega_k^{\parallel} t -\eta_\k^{\parallel})\ ,
\eeq
and similarly for the three transverse chiral components
$\bold{\dphi}_\perp(\r,t)$.
Here the energy $\eps_k=\hbar\omega_k$ is determined by
the Klein-Gordon dispersion relation, $\eps_k^2=\hbar^2c^2 k^2+\mu^2c^4$
(with $\mu$ being the appropriate effective mass, $\mup$ or $\mut$).
The phase $\eta_\k$ is random in the interval $(0,2\pi)$
and is thus trivial to sample.
Furthermore,
the real (and positive) amplitude $C_\k$
can be related to the number of quanta $n_\k$ 
by considering the energy carried by the mode,
\beq\label{nk}
E_\k=n_\k \eps_k={\eps_k^2\over\hbar^3c^3} C_\k^2\,\, 
\Rightarrow\,\,
C_\k = \left[\hbar^3c^3\ {n_\k \over \eps_k}\right]^{1\over2}\ .
\eeq
We have here omitted the zero-point energy \cite{Kapusta},
thereby eliminating the associated ultraviolet divergence.
Although not quite correct,
this approach is justified {\em a posteriori} by the apparent good quality of
the resulting approximate treatment.
(The dynamical tests discussed in sect.\ \ref{dynamics} are here very important,
since the statistical samples would not remain stationary under time evolution
if the treatment were substantially wrong.)

Thus the problem has been reduced to sampling the number of quanta $n_\k$.
Since the probability for finding a particular number of quanta
in the mode $\k$ is given by
$P(n_\k)=(1-\exp(-\eps_k/T))\exp(- n_\k\eps_k/T)$,
it is elementary to sample the integer $n_\k$ appropriately.\footnote{The
corresponding algorithm for this task is very simple
because $n_\k$ can be regarded as counting the number of successive times
the sampling of a standard random number (\ie\ uniform on $(0,1)$)
yields a value below $\exp(-\eps_k/T)$.}
We note that the thermal average of $n_\k$
is equal to the occupancy $f_k$ employed in the calculation of the entropy,
$\prec n_\k \succ =  f_k$.

Once the amplitudes and phases have been selected,
the expansion (\ref{dphi}) readily yields the initial value of
the field fluctuations, $\dpphi(\r,0)$.
The corresponding conjugate momentum, $\delta\ppsi\equiv\hbar\del_t\dpphi$,
readily follows since (\ref{dphi}) implies
\beq\label{dpsi}
\delta\psi_\parallel(\r,t) =\left(2\over\Omega\right)^{1\over2}{\sum_\k}
\eps_{k}^\parallel C_{\k}^\parallel\ \sin(\k\cdot\r-\omega_{k}^\parallel t 
-\eta_{\k}^\parallel)\ .
\eeq
The entire state of the chiral field, $(\pphi(\r),\ppsi(\r))$,
has then been selected at the time $t=0$.
When the equations of motion are propagated by a leap-frog method,
the field strength $\dpphi(\r)$ is calculated at the times $t_n=n\Delta t$
while the momentum $\delta\ppsi(\r)$ is obtained at the intermediate times.
The appropriate initial $\delta\ppsi(\r)$ can then easily be obtained
by evaluating the expansion (\ref{dpsi}) at $t=\Delta t/2$,
after $C_\k$ and $\eta_\k$ have been selected.

Finally,
since the sampling has been done in a system aligned with
the order parameter $\ul{\pphi}$ (in which the mass tensor is diagonal),
an $O(4)$ rotation of the sampled field configuration is required
in order to express the state with respect to the chiral directions.
This is readily accomplished on the basis of the angles
$(\chi_0,\vartheta_0,\varphi_0)$ specifying the direction of $\ul{\pphi}$.

\subsubsection{Occupation numbers}
\label{occup}

Since the quasi-particles represent bosonic modes,
it is useful to know how large the occupancies can become.
The thermal occupation numbers are given by
$\prec n_\k \succ=f_k=1/(\exp(\eps_k/T)-1)$,
and since $\mut<\mup$ in any given scenario
the largest occupancies occur for the transverse modes,
$f_k^\perp>f_k^\parallel$.
Moreover,
for a given temperature $T$,
$f_k$ is largest when the momentum vanishes, $f_0=1/(\exp(\mu c^2/T)-1)$,
which then provides an upper bound on $f_k$
(recall that $k>0$ for the quasi-particles modes).

Considered as functions of $T$,
the bounds $f_0^\parallel$ and $f_0^\perp$ start out from zero,
display maxima well above $T_c$,
and then drop off towards a common constant value at high temperatures.
The (common) limiting occupancy is $f_0\approx0.26$
since $\mu c^2\approx1.59T$ when $T\to\infty$.
(This feature is a direct consequence of the repulsive self-interaction
of the chiral field
and is in marked contrast to the ever increasing occupancy
characteristic of free bosons in a thermal bath.)
The maximum values attained by $f_0^\parallel$ and $f_0^\perp$ 
are about 0.38 and 0.48
and occur at approximately a temperature of 265 and 235 MeV, respectively.
Since these values are well below unity,
the system is never very degenerate.

Nevertheless, there is an important advantage to using Bose-Einstein
rather than Boltzmann statistics for the field fluctuations,
as we will now discuss.
When classical statistics is used,
the occupation probability is $n_k\approx T/\eps_k$ when $\eps_k\gg T$
and therefore the total quasi-particle density grows
as the square of the upper limit on $\eps_k$,
a manifestation of the Rayleigh-Jeans divergence.
Such a description would be entirely wrong in the present context and,
moreover, it would be numerically ill-behaved.
By contrast, the quantal occupation probability falls off exponentially
and the density of quanta is finite.
By adopting the Bose-Einstein treatment,
we then get a much more realistic formal description of the statistical
properties of the system and,
in addition, the numerical treatment becomes straightforward.
Of course, if a configuration sampled on the basis of the quantal statistics
is propagated for a sufficient length of time,
it will eventually exhibit classical features,
since the equation of motion is entirely classical.
It is therefore fortunate that our treatment is expected to be applied only
to processes that are far faster
than the time scales assoicated with the reversion to classical statistics.
In particular,
for the formation of disoriented chiral condensates in high-energy collisions
the relevant time scales are of the order of a few $\fm/c$,
while our numerical studies exhibit no ultraviolet run-away
for at least several tens of $\fm/c$.

\subsection{Correlation function}

It is particularly interesting to calculate the correlation function
of the chiral field
since this quantity determines the spectral distribution
of the emitted field quanta.

The density matrix associated with the quasi-particle degrees of freedom
is a $4\times4$ tensor,
\beq\label{C}
\bold{C}(r_{12},t_{12})\ \equiv\
\prec\dpphi(\r_1,t_1)\ \dpphi(\r_2,t_2)\succ\ .
\eeq
The average is over the ensemble of systems considered,
in the present case a thermal ensemble held at the temperature $T$.
Since a system in equilibrium is invariant in time,
the correlation function depends only on the time difference
$t_{12}=t_1-t_2$.
Analogously,
the translational symmetry of the scenario
implies that the spatial dependence is via the separation $\r_{12}=\r_1-\r_2$.
Moreover,
to the extent that there is invariance under spatial rotations,
only the magnitude $r_{12}=|\r_{12}|$ enters.
In principle, these spatial symmetries are broken
when a finite box is considered
but the effect is insignificant and can be disregarded in the present study.

Utilizing the expansion (\ref{dphi}),
it is elementary to show that the correlation tensor $\bold{C}$ is diagonal
with the elements $C_\parallel$ and $C_\perp$, where
\beq\label{Cp}
C_\parallel(r,t)\ =\
{\hbar^3c^3\over\Omega}{\sum_\k}' {1\over\eps_k}
{\cos(\k\cdot\r-\omega_k t) \over \rme^{\beta\eps_k}-1}\
\asymp\ {1\over2\pi^2}{\hbar c\over r}
\int_{\mup c^2}^\infty d\eps\
{\sin{kr} \over \rme^{\beta\eps}-1} \cos\omega_k t\ ,
\eeq
with $\eps_k^2=\hbar^2\omega_k^2=\hbar^2k^2c^2+\mup^2c^4$.
An analogous expression holds for $C_\perp(\r,t)$.
Since $\hbar^3c^3 f_k/\eps_k$ is equal to
the thermal average $\prec A_\k^2 \succ$,
we recognize that the result (\ref{Cp})
corresponds to the expression (146.10) given in ref.\ \cite{LL}.

The usual correlation function is the trace of $\bold{C}$,
\beq\label{C12}
C\ \equiv\ \prec\dpphi(\r_1,t_1) \cdot \dpphi(\r_2,t_2)\succ\
=\ {\rm tr}\ \bold{C}\
=\ C_\parallel(r_{12},t_{12}) +3C_\perp(r_{12},t_{12})\ .
\eeq
Its overall magnitude is set by its value at zero,
which is simply the corresponding fluctuation in the field strength,
$C(0,0)=\prec \delta\phi^2 \succ$.
It is therefore convenient to divide $C$ by this quantity
and so define the reduced function
$C_{12}\equiv C(r_{12},t_{12})/\prec\dphi^2\succ$.
Since $C_{12}$ is unity when $\r_1=\r_2$ and $t_1=t_2$,
it expresses the space-time attenuation of the correlation
between the field-strength fluctuations at different space-time points.
In general, $C_{12}\sim (1/r_{12})\exp(-\mu cr_{12}/\hbar)$
in the limit of large separations, $r_{12}\to\infty$,
so that $\hbar/\mu c$ provides a simple measure of the correlation length.
In the special case when the effective mass vanishes,
the reduced equal-time correlation function is given on analytical form,
$C_{12} \asymp (3/\zeta)(\coth\zeta -1/\zeta)$,
where $\zeta\equiv\pi T r_{12}/\hbar c$.
In this extreme case,
the correlation function falls off only as $\sim1/r_{12}$.

Figure \ref{f:C12-Tc} shows the reduced correlation function $C_{12}$
obtained at the critical temperature, $T=T_c$.
Its appearance depends on the magnitude of the order parameter, $\phi_0$,
through the effective masses.
As $\phi_0$ is increased from zero
to its vacuum value $f_\pi=92\ \MeV$,
the corresponding effective masses increase from zero to their free values
and the correlation function falls off ever more rapidly.
The correlation function thus exhibits a significant sensitivity
to the order parameter.
We note in particular that for the most probable value,
$\phi_0\approx87\ \MeV$,
the attenuation of $C_{12}$ is considerably faster than for $\phi_0=0$
and its width has dropped by over a factor of two.
It is therefore important to take proper account of the order parameter
when calculating the correlation function.

Figure \ref{f:C12} shows how the reduced correlation function evolves
with temperature
when the most probable magnitude of the order parameter is employed.
At high temperatures the effective masses grow
nearly in proportion to $T$
so then the correlation length tends to zero,
as is borne out by the steady shrinking of $C_{12}$.
For temperatures below critical,
the field fluctuations are predominantly associated with the transverse modes,
since those have the smallest effective mass, $\mut\ll\mup$,
and the correlation length grows ever larger.

It is common to characterize the system by the ``correlation length'',
defined as the first moment of the equal-time correlation function
which diverges when $\mu\to0$.
For the present discussion,
it is more convenient to characterize $C_{12}$
by its full width at half maximum for equal times, $\Gamma_{12}$,
since this quantity is always easy to extract,
even when the mass vanishes.
Figure \ref{f:FWHM} shows this measure of the correlation length
as a function of temperature,
using the most probable value of $\phi_0$.
For temperatures near and below $T\approx200\ \MeV$
the dominant fluctuations are perpendicular to the order parameter
since the corresponding effective mass is relatively small.
For higher temperatures the asymptotic regime is approached
where the chiral symmetry is approximately restored and
the fluctuations are similar in all four chiral directions.

It is sometimes of interest to also consider correlation functions
involving  $\dppsi$,
the time derivative of the local field strength \cite{Kluger}.
Those can be obtained in a similar manner.

\subsection{Radiation spectra}
\label{radiation}

The evolving chiral field may give rise to mesonic radiation,
in analogy with the emission of photons
by a time dependent electromagnetic field.
The rate for production of a field quantum
having energy $E$ and momentum $\p$
is proportional to the square of the corresponding Fourier amplitude
\cite{Horn,GKW:PRC20,Amado:PRD51},
\beq
E{dN_j\over d\p}\ \sim\ \left|
\int d\r\int dt\ \dphi_j(\r,t)\ \rme^{-{i\over\hbar}(\p\cdot\r-Et)} \right|^2\ .
\eeq
Here $j$ denotes the particular $O(4)$ direction considered
so that $j=0$ gives rise to isoscalar $\sigma$-like mesons and
$j>0$ represent three components of the isovector pion-like mesons
(with $j=3$ corresponding to $\pi_0$, say).

A uniform system in equilibrium has both temporal and spatial invariance
and the specific radiation rate, $\nu_j(\p,E)$
(\ie\ the production per unit volume and per unit time),
is then given by the Fourier transform of the correlation function
\cite{Gavin:QM95},
\beq
\nu_j(\p,E)\ \equiv\ {1\over\Omega t_0} \prec E{dN_j\over d\p}\succ\
\sim\ \int d\r \int dt\ C_j(r,t)\ \rme^{{i\over\hbar}(\p\cdot\r-Et)}\
\equiv\ {\cal C}_j(\p,E)\ ,
\eeq
where $\Omega$ is the volume of the system
and $t_0$ is the time interval considered.
The Fourier transform of the correlation function
is easy to obtain in the continuum limit,
\beq
{\cal C}_j(\p,E)\ \asymp\
\hbar^3c^3 {\pi \over \rme^{\beta E}-1}\ \delta(m^2c^4 - \mu_j^2c^4)\ ,
\eeq
where we have assumed that the energy and momentum of the radiated mesons
are related by $E^2=p^2c^2+m^2c^4$.

The above result holds for a specified value of the order parameter $\phi_0$
which determines the effective mass $\mu_j$.
As we have discussed,
the order parameter has in general a statistical distribution, $P(\phi_0)$,
giving rise to corresponding distributions of the effective masses,
$P_\parallel(\mu)$ and $P_\perp(\mu)$ (see sect.\ \ref{mass-dist}).
The resulting production rate can then be obtained by integrating over
the appropriate thermal mass distribution.
Thus, for example,
the specific production rate for $\pi_0$ mesons is
\beq\label{nu0}
\nu_{\pi^0}(\p,E) \sim \int d\mu^2c^4\ P_\perp(\mu^2c^4)\ {\cal C}_\perp(\p,E)
=\hbar^3c^3 {\pi P_\perp(m_\pi^2c^4) \over \rme^{\beta E}-1}\ ,
\eeq
and the rates for the charged pions are similar.\footnote{The rates
considered here pertain to the idealized scenario
of a macroscopically uniform system (possibly enclosed in a periodic box),
and hence they differ from those describing the emission from a finite source
into the surrounding vacuum.
In particular,
the familiar kinematic enhancement of the faster-moving ejectiles is absent.}
This result is easy to interpret:
The spectral distribution of the radiated mesons
is of thermal Bose-Einstein form,
with the temperature given by the value characterizing the source itself,
and the overall normalization of the radiation rate
is proportional to the probability
that the space and time evolution of the field
matches the particular dispersion relation for the type of meson considered.

As an illustration,
fig.\ \ref{f:pions} displays the strength function $P_\perp(m_\pi^2c^4)$
obtained by evaluating eq.\ (\ref{Pmu}) for $\mu=m_\pi$
as a function of the source temperature.
As the temperature is decreased towards zero,
the centroid of the $\mut$ distribution moves down
towards the free value $m_\pi$
and at the same time its width keeps shrinking towards zero
(see fig.\ \ref{f:Tmu}).
Through the supercritical regime
the pion strength then grows approximately exponentially.
Then a plateau is reached
where the increase caused by the approach of the centroid to $m_\pi$
is counterbalanced by the decrease due to the shrinkage.
As a result,
the strength is nearly constant from 90 to 30 MeV.
Finally, after the centroid has practically reached $m_\pi$,
the free strength exhibits a rapid rise
as the $\mut$ distribution approaches a $\delta$ function.
On the basis of this result,
one would expect radiation of free pions to be unimportant
until the temperature has dropped below $T_c$.

\section{Dynamical tests}
\label{dynamics}
Once the initial state has been prepared,
for example by means of the statistical sampling described above,
the chiral field may be propagated in time
by means of the equation of motion (\ref{EoM}),
which is straightforward to implement in either $x$ or $k$ representation.

It is possible to exploit the dynamical evolution to test the validity
of our approximate statistical treatment.
If the system is ergodic,
as would be expected because of its non-linear interaction,
a dynamical trajectory will explore the space of possible field configurations
in accordance with the appropriate microcanonical weight.
Conversely,
an ensemble of field configurations that has been sampled statistically
should not exhibit any change under time evolution.
These features provide a convenient means for checking our treatment
and we give two illustrations below.\footnote{It is important to recognize
that whereas the statistical properties have been obtained
by assuming that the quasi-particle degrees of freedom
are effectively decoupled,
no such assumption is being made in the dynamics,
since the trajectories are obtained by solving the
full equation of motion (\ref{EoM}).}

\subsection{Average field strength}
\label{mean}
Perhaps most vividly,
we show in fig.\ \ref{f:dyn0} the early trajectories of the order parameter
$\ul{\pphi}$ for a sample of eight configurations,
considering a box with $L=8\ \fm$ and either $T=200\ \MeV$ or $T=240\ \MeV$.
The dashed contours are those already given in fig.\ \ref{f:contour}
indicating where the projected probability density has fallen to half
its maximum value and the centroids are indicated as well (open dots).
For each individual trajectory,
the initial location is indicated by the solid dot
and the attached solid curve shows the trajectory
up to $t=1\ \fm/c$.
The fact that the initial points reflect the
calculated statistical distribution provides
an elementary test of the numerical sampling algorithms.
Less trivial is the fact that the dynamical trajectories indeed appear
to explore the region predicted by the approximate statistical distribution.

This correspondence can be made more quantitative
by studying how the distribution of the order parameter
evolves in the course of time.
This analysis is illustrated in fig.\ \ref{f:dynphi0}.
The approximate distribution $P_\phi(\phi_0)$ given in (\ref{Pphi})
is indicated by the solid curves for $T=180$ and $T=240$ MeV.
Forty individual systems have then been prepared by sampling
their field configurations as described above
and they have subsequently been propagated by the equation of motion (\ref{EoM})
up to the time $t=10\ \fm/c$.
In the course of the evolution,
the value of $\phi_0$ is extracted at regular intervals
and binned into slots that are 5 MeV wide.
In this manner the time-averaged distribution of $\phi_0$ can be determined
and the dashed curves display the result
(which is not sensitive to
an increase of either the maximum time or the sample size).
The overall agreement with the initial distribution is very good.
There is generally is slight shift outwards,
amounting typically to 1-2 MeV,
which suggests that our approximate thermal distributions
are centered at correspondingly too low values of $\phi_0$.
Consequently,
for applications where such a relatively minor displacement is unimportant,
the test demonstrates the validity of our approximate method
for obtaining the statistical properties of the model.

\subsection{Field fluctuations}
\label{flucts}
Figure \ref{f:dynC} displays the correlation function
for the pion components, $C_{12}^\pi(s_{12})$,
for a box with $L=5\ \fm$ prepared at $T=240\ \MeV$.
The short-dashed curves show the correlation function
obtained on the basis of ten sampled configurations,
and the long-dashed curves then indicate the corresponding result
after those systems have been evolved up to the time $t=10\ \fm/c$.
The two curves that go up again have been obtained by
aligning the relative separation $\r_{12}$ along one of the cartesian axes,
while it is directed diagonally for the other two curves
(the periodicity is then $\sqrt{3}L$
and so their eventual rise is only barely visible).
For reference is shown the exact thermal correlation function
for either the finite box considered (solid curves)
or the continuum limit (dots).
While this latter curve tends to zero for large separations
(and in fact falls off monotonically),
the correlation function for a finite box drops to a negative value,
because its spatial average must vanish.

The correlation function remains remarkably invariant in the course of time.
This indicates that our treatment, including the sampling procedure,
in fact yields a good approximation to the correlated field fluctuations.
As a quantitative measure,
one may consider the full width at half maximum $\Gamma_{12}$.
The continuum value is 1.33 fm,
slightly larger than the thermal result for the finite box, 1.30.
For the sample of ten initial configurations we find 1.256 and 1.271
for the cartesian and diagonal directions, respectively,
which have evolved into 1.263 and 1.290 at $t=10\ \fm/c$.
So there is no significant change in the width of the correlation function.

\section{Concluding remarks}
\label{conclusion}

The present work was motivated by the current interest
in disoriented chiral condensates,
particularly the various dynamical simulations
carried out with the linear $\sigma$ model
\cite{Rajagopal:NPB404,GGP93,Kluger,Huang:PRD49,Asakawa,Csernai}.
Those calculations follow the non-equilibrium
evolution of the cooling chiral field
in order to ascertain the degree to which coherent domains develop.
Since the dynamics is inherently unstable,
with the low-momentum modes experiencing rapid amplification,
one may expect a significant sensitivity of the results
to the initial conditions,
with a commensurate degree of difficulty regarding their interpretation.
Consequently,
caution is required
when characterizing the ensemble of initial field configurations employed.

In order to provide a useful framework for this aspect of the problem,
we have explored the statistical properties of the linear $\sigma$ model,
by confining the system to a rectangular box held at a fixed temperature.
Although this problem can be treated exactly \cite{Bernstein},
we have found it preferable to linearize the equations of motion
since our view is towards practical calculations.
The resulting treatment then becomes very simple
and appears to be sufficiently accurate in the present context. 

The problem separates into one concerning the spatial average of
the field, the order parameter,
and another dealing with the field fluctuations,
referred to as the quasi-particle degrees of freedom.
The latter are described approximately in terms of effective masses
that depend on both the order parameter and the temperature,
but are independent of the symmetry breaking $H$ term;
these were presented in fig.\ \ref{f:mu}.

The partition function then takes on a corresponding separable form and,
as a consequence,
it is possible to develop a simple method
for performing a statistical sampling of the thermal equilibrium
field configurations, including their time derivatives,
at any temperature.
This result, having a clear physical basis,
is expected to be directly useful as practical means for initializing
the dynamical simulations of the chiral field 
of the type carried out recently by several groups
\cite{Rajagopal:NPB404,GGP93,Kluger,Huang:PRD49,Asakawa,Csernai},
thus making it easier to interpret the numerical results.

Moreover,
our specific illustrations provide useful insight
into the equilibrium properties.
In particular,
it appears to be unrealistic to start the order parameter off
with a value equal to zero.
Indeed,
at the critical temperature the most probable order parameter
is much closer to its vacuum value $f_\pi$ than to zero.
The relationship between temperature and order parameter
was summarized in fig.\ \ref{f:Tphi}
and a more global impression of the distribution of the the order parameter
(including its degree of misalignment)
can be gained from the contour plots in fig.\ \ref{f:contour}.

Since the order parameter is thus very unlikely to vanish,
the effective quasi-particle masses remain finite.
Consequently,
the statistical equilibrium distribution remains well behaved
at all temperatures
and the change from the ``restored'' phase to the normal one is fairly gradual.
However, the finite size generally reduces the effective masses,
thereby bringing the system closer to criticality.

Of course,
the statistical properties are of most practical interest
at the relatively high temperatures characteristic of the initial stage
of the high-energy collision.
Once the chiral field has been initialized accordingly,
any instabilities and associated amplifications
will be automatically included in the dynamical propagation
and the system can generally be expected to quickly move out of equilibrium.
The equilibrium results can then provide a meaningful reference
against which to analyze the deviation from equilibrium
at any stage in the dynamical relaxation process.

Additionally,
we illustrated briefly the equilibrium form of the correlation function
which is an object of primary interest.
Indeed,
it is the correlation length that properly expresses the ``domain size''
governing the conjectured anomalous pion radiation.
Essentially,
what one would expect to see at the end
is a stretched version of the initial correlation function
since the long wave lengths are the most unstable
and so will contribute in an ever larger propertion.
This underscores the importance of starting out
with chiral fields having physically reasonable correlation properties.
To illustrate the use of the correlation function,
we derived the rate at which real pion mesons are created by th field,
and subsequently we calculated the dependence of the free pion strength
on the temperature of the system.

Finally,
we sought to assess the degree of validity of our approach
by subjecting sampled field configurations to the exact time evolution.
This convenient means of testing suggested that
the approximate treatment is of sufficient accuracy to be of practical utility.
We therefore anticipate that it may find use in simulation studies,
such as those exploring disoriented chiral condensates
in high-energy collisions.
~\\

The author is pleased to acknowledge helpful discussions with many colleagues,
including
P.\ Bedaque, L.P. Csernai, S.\ Gavin, J.I.\ Kapusta, Y.\ Kluger,
V.\ Koch, L.\ McLerran, R.D.\ Pisarski, R.\ Vogt, and X.N.\ Wang.


\newpage
\bfig
\vspace{1in}\hspace{-0.5in}
\rotateright{\psfig{figure=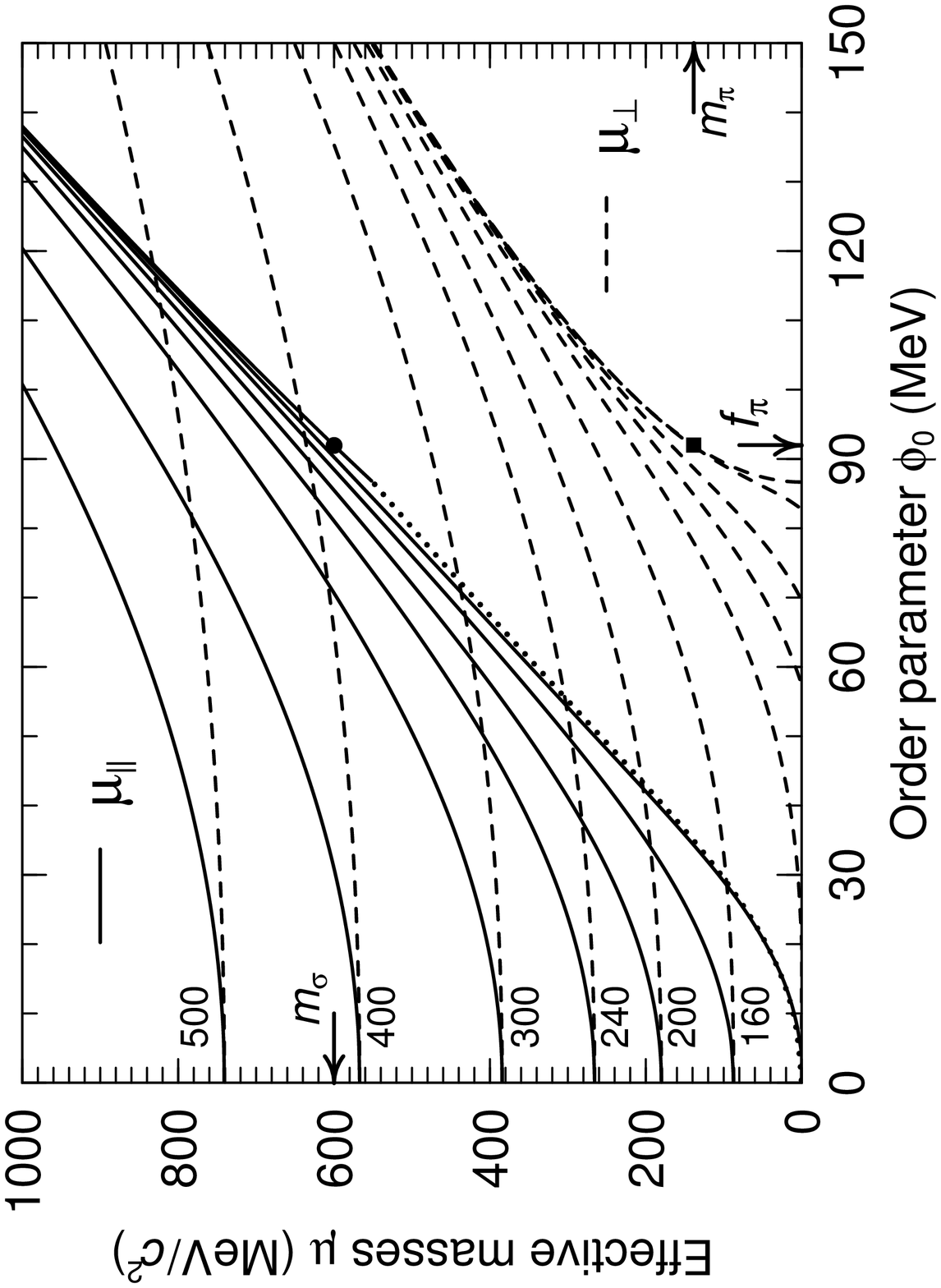,width=6in,height=4.8in}}
\vspace{-1in}
\caption{Effective masses.}
\label{f:mu}
The effective masses $\mup$ (solid) and $\mut$ (dashed),
as functions of the magnitude of the order parameter, $\phi_0$,
for a range of temperatures:
$T$ = 0, 40, 80, 100, 122.63 (=$T_c$), 160, 200, 240, 300, 400, 500 $\MeV$,
calculated in the thermodynamic limit where the box size is large, $L\to\infty$.
For a temperature above $T_c$,
the two effective-mass curves start out at $\phi_0=0$ with degenerate values,
whereas below $T_c$
they only exist if $\phi_0$ is sufficiently large.
The corresponding starting points for $\mup$    
are connected by the dotted curve and,
since $\mup$ is then nearly independent of $T$,
only the curve for $T=0$ has been shown.
The vertical arrow points to the vacuum value of the order parameter,
$\phi_{\rm vac}=f_\pi=92\ \MeV$,
and the free mass values $\mup=m_\sigma=600\ \MeV/c^2$
and $\mut=m_\pi=138\ \MeV/c^2$
are indicated by the horizontal arrows.
The locations of the corresponding points in the diagram
are shown by the two solid symbols.
\efig
\newpage
\bfig
\vspace{1in}\hspace{-0.5in}
\rotateright{\psfig{figure=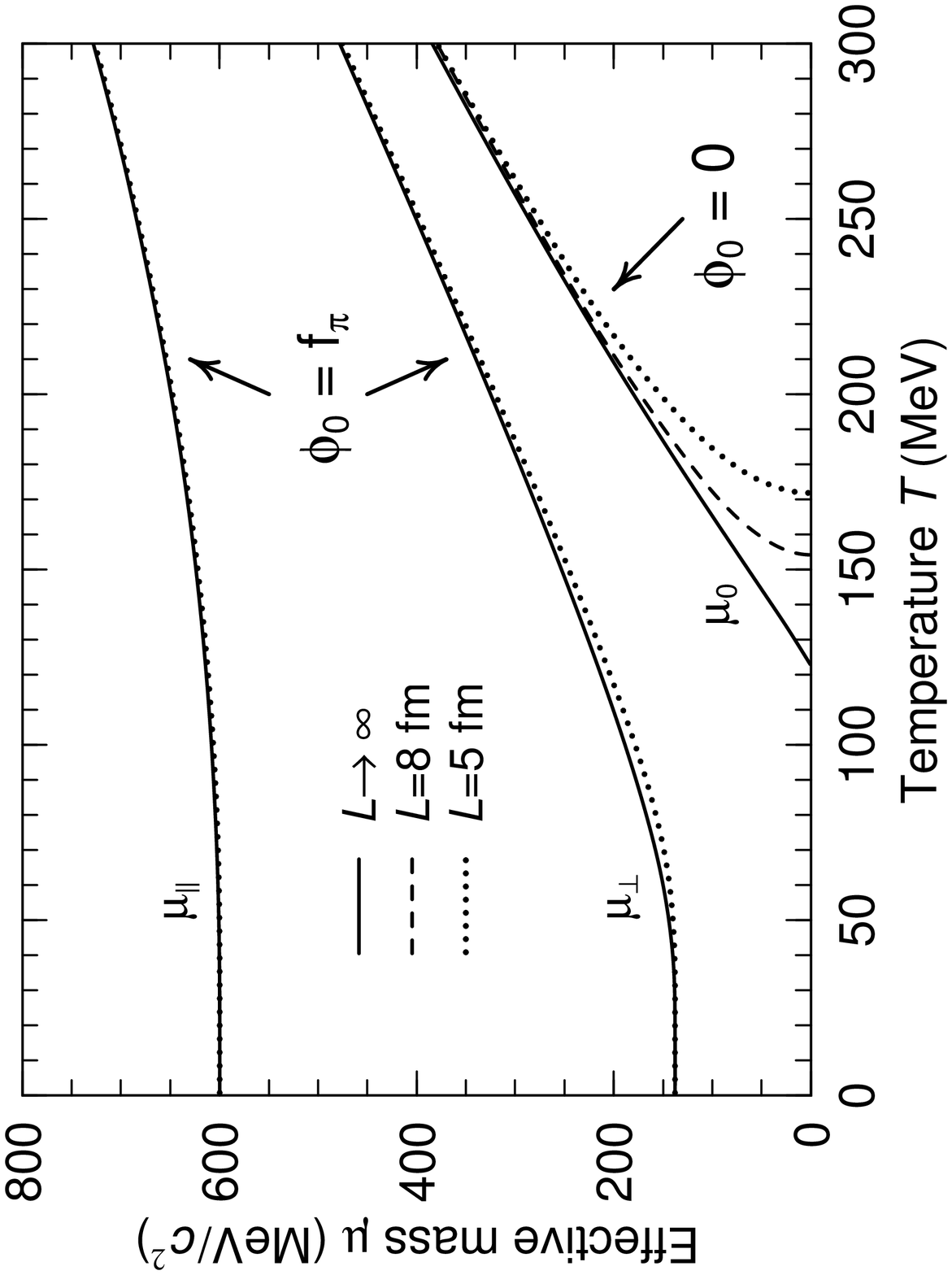,width=6in,height=4.8in}}
\vspace{-1in}
\caption{Effect of finite size on the effective masses.}
\label{f:muL}
The effective masses are shown as functions of the temperature
for either $\phi_0=0$, when the $O(4)$ symmetry is restored,
or for $\phi_0=f_\pi$, the physical vacuum value. 
Three different scenarios have been considered:
either the thermodynamic limit of large volume 
in which the quasi-particle spectrum is continuos (solid curve),
corresponding to the scenario of fig.\ \protect\ref{f:mu},
or a finite cubic box with side length
$L=8\ \fm$ (dashed) or $L=5\ \fm$ (dotted),
where the quasi-particle modes are quantized.
\efig
\bfig
\vspace{1in}\hspace{-0.5in}
\rotateright{\psfig{figure=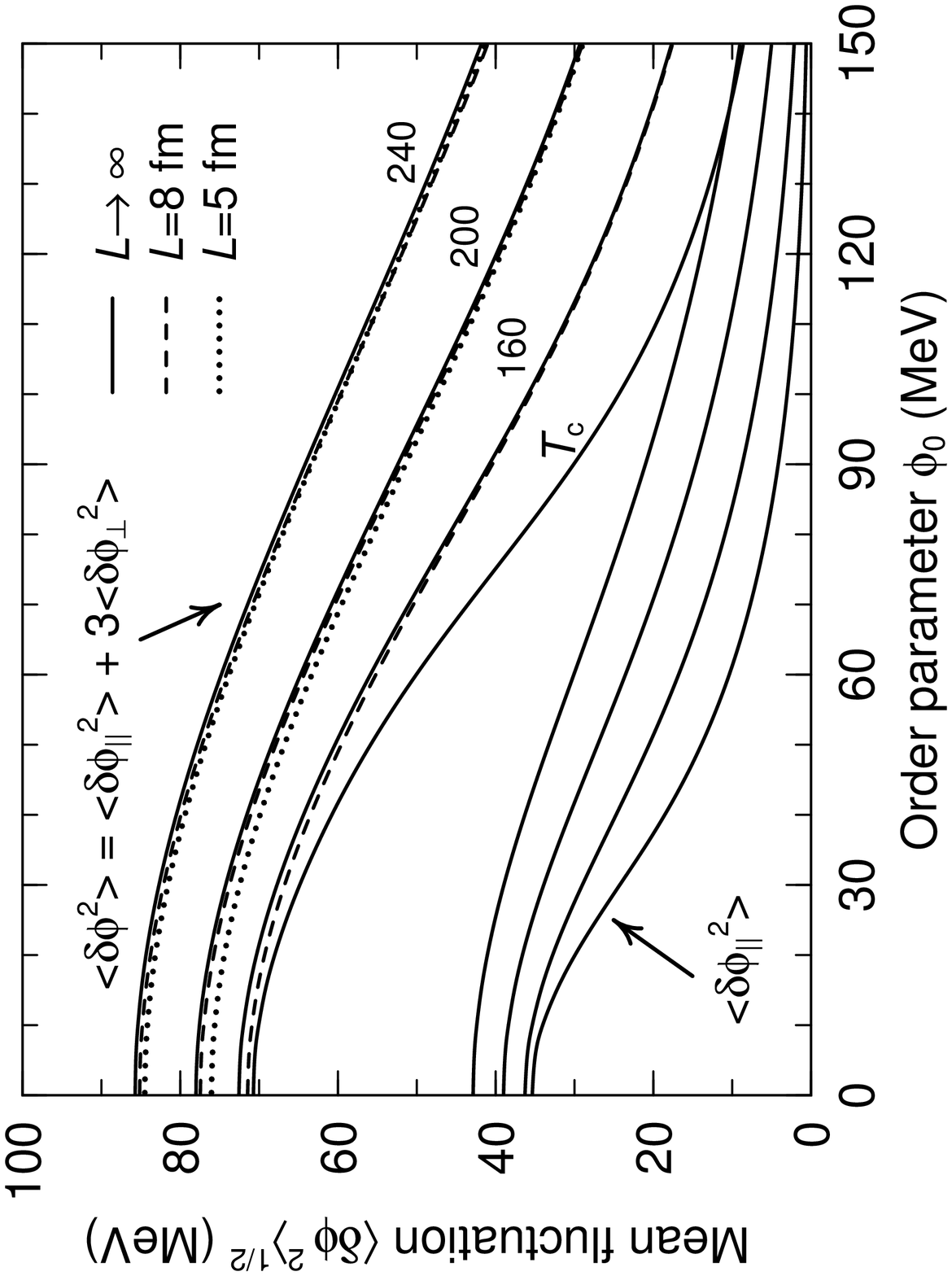,width=6in,height=4.8in}}
\vspace{-1in}
\caption{Fluctuations.}
\label{f:phi2}
The typical magnitude of the spatial fluctuations
in the field strength,  
$\prec\dphi^2\succ^{1/2}$,
as a function of the magnitude of the order parameter, $\phi_0$,
for specified temperatures $T$: 122.63 (=$T_c$), 160, 200, 240 MeV.
The system is enclosed in a cubic box of side length
$L\to\infty$ (solid curves), $L=8\ \fm$ (dashed), or $L=5\ \fm$ (dotted).
The lower four curves display the fluctuations
along the chiral direction of the order parameter $\ul{\pphi}$,
while the upper curves represent the total fluctuation.
\efig
\bfig
\vspace{1in}\hspace{-0.5in}
\rotateright{\psfig{figure=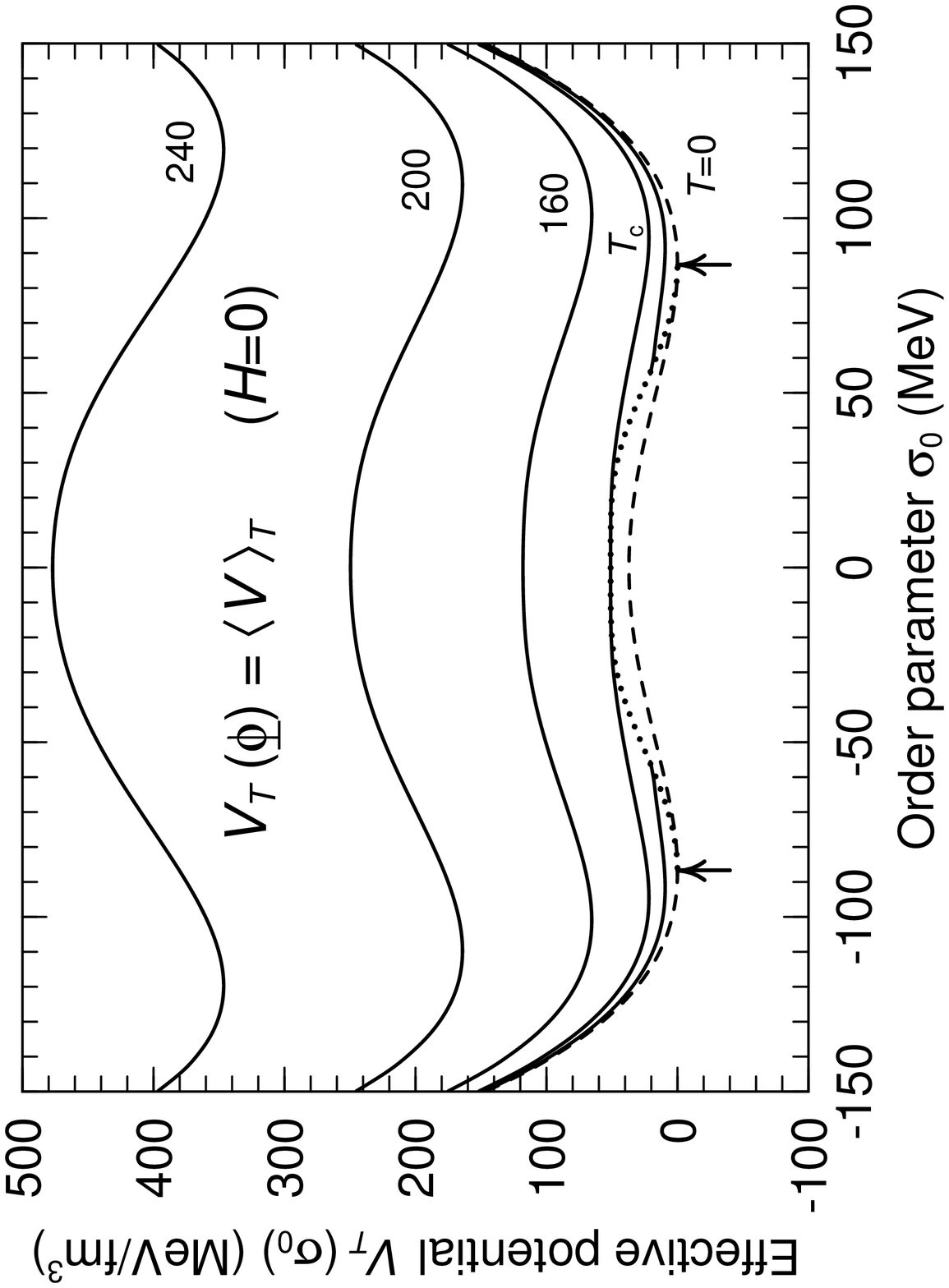,width=6in,height=4.8in}}
\vspace{-1in}
\caption{Effective potential (part $a$: $H=0$).}
\label{f:V0}
The effective potential energy density $V_T$ along the $\sigma$ axis
for either the $O(4)$ symmetric model having $H=0$ (part $a$)
or with the adopted positive value of $H$ (part $b$).
The solid curves show the results for a number of temperatures:
100, 122.63 (=$T_c$), 160, 200, 240 MeV.
For $T<T_c$ the effective potential curve
starts at a certain minimum value of $\phi_0$ between 0 and $v$,
These starting points are connected by the dotted curve,
while the dashed curve shows the bare potential $V_0$
obtained when fluctuations are neglected.
The arrows point to the minima of $V_T$;
for $H>0$ there is only a single minimum (located at $\sigma_0=f_\pi$),
while for $H=0$ the degenerate ground-state minima
form the surface of the 4-sphere determined by $\phi=v$.
The value of the effective potential
corresponding to other orientations of the order parameter
can be easily obtained by noting that
the directional dependence of $V_T$
is given by $-H\phi_0\cos\chi_0/\hbar^3c^3$.
\efig
\addtocounter{figure}{-1}
\bfig
\vspace{1in}\hspace{-0.5in}
\rotateright{\psfig{figure=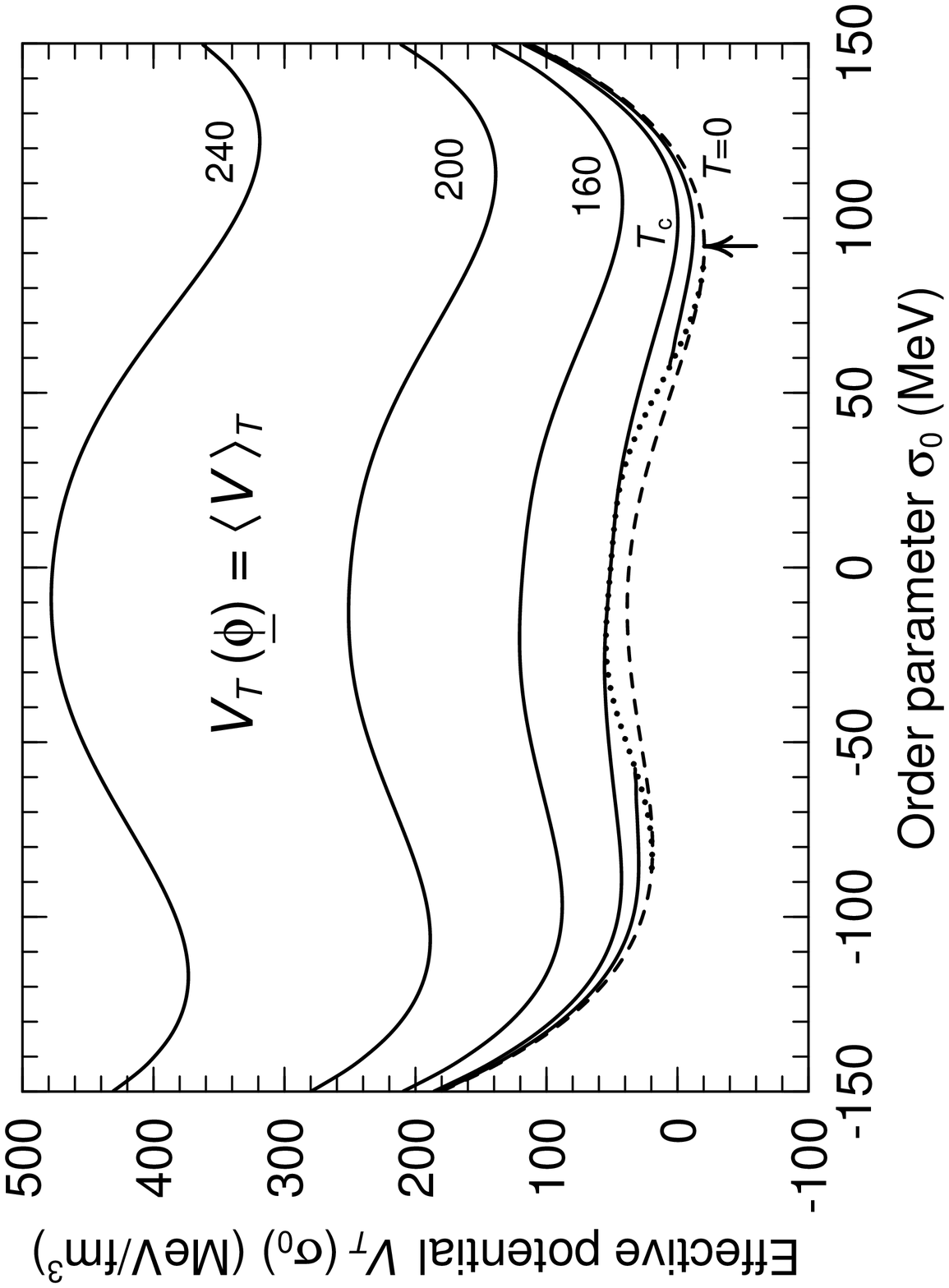,width=6in,height=4.8in}}
\vspace{-1in}
\caption{Effective potential (part $b$: $H>0$).}
\label{f:V}
The effective potential energy density $V_T$ along the $\sigma$ axis
for either the $O(4)$ symmetric model having $H=0$ (part $a$)
or with the adopted positive value of $H$ (part $b$).
The solid curves show the results for a number of temperatures:
100, 122.63 (=$T_c$), 160, 200, 240 MeV.
For $T<T_c$ the effective potential curve
starts at a certain minimum value of $\phi_0$ between 0 and $v$,
These starting points are connected by the dotted curve,
while the dashed curve shows the bare potential $V_0$
obtained when fluctuations are neglected.
The arrows point to the minima of $V_T$;
for $H>0$ there is only a single minimum (located at $\sigma_0=f_\pi$),
while for $H=0$ the degenerate ground-state minima
form the surface of the 4-sphere determined by $\phi=v$.
The value of the effective potential
corresponding to other orientations of the order parameter
can be easily obtained by noting that
the directional dependence of $V_T$
is given by $-H\phi_0\cos\chi_0/\hbar^3c^3$.
\efig
\bfig
\vspace{1in}\hspace{-0.5in}
\rotateright{\psfig{figure=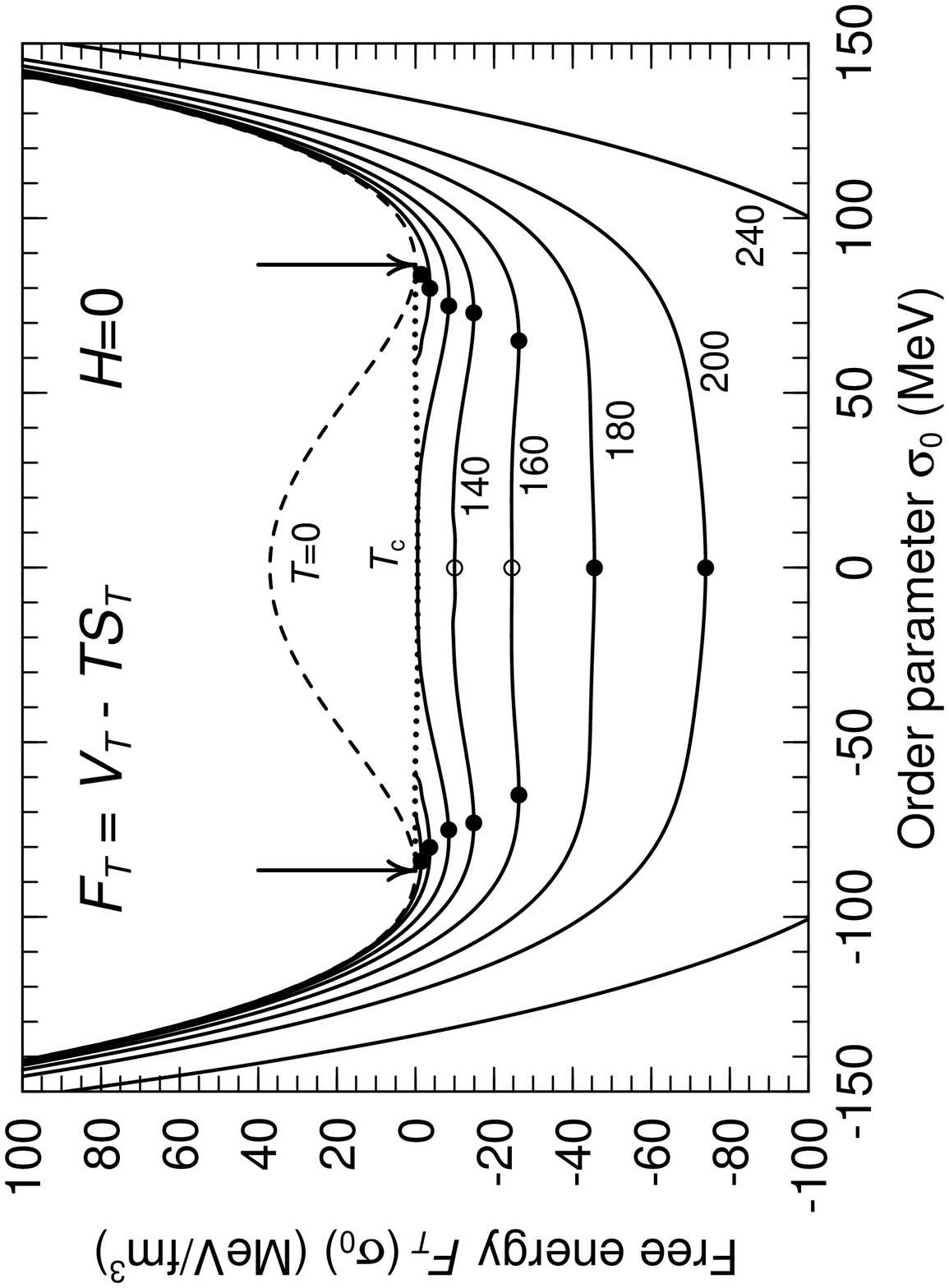,width=6in,height=4.8in}}
\vspace{-1in}
\caption{Free energy (part $a$: $H=0$).}
\label{f:F0}
The free energy density $F_T(\ul{\phi})$ along the $\sigma$ axis
for either the $O(4)$ symmetric model having $H=0$ (part $a$)
or with the adopted positive value of $H$ (part $b$).
The solid curves show the results for a number of temperatures.
For $T<T_c$ the curve
starts at a certain minimum value of the magnitude $\phi_0$
and these starting points are connected by the dotted curve,
while the dashed curve shows the result
obtained when the temperature is neglected.
The arrows point to the minima of $F_T$;
for $H>0$ there is only a single minimum,
while for $H=0$ the degenerate minima form the surface of a 4-sphere.
For each temperature the location of the minima are indicated by the solid dots.
(For $H=0$ the shallow secondary minima at $\phi_0=0$ are shown by open dots.)
\efig
\addtocounter{figure}{-1}
\bfig
\vspace{1in}\hspace{-0.5in}
\rotateright{\psfig{figure=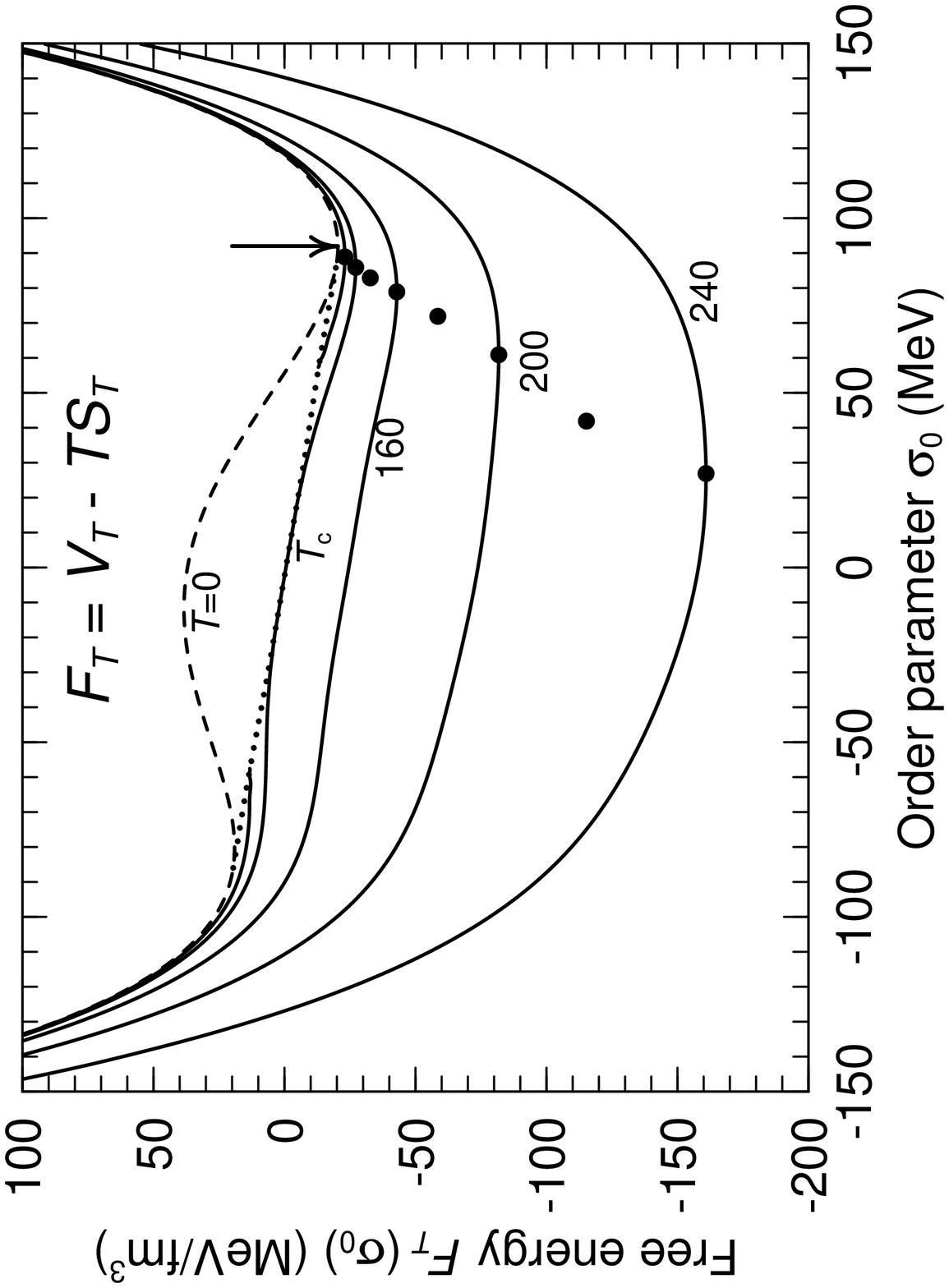,width=6in,height=4.8in}}
\vspace{-1in}
\caption{Free energy (part $b$: $H>0$).}
\label{f:F}
The free energy density $F_T(\ul{\phi})$ along the $\sigma$ axis
for either the $O(4)$ symmetric model having $H=0$ (part $a$)
or with the adopted positive value of $H$ (part $b$).
The solid curves show the results for a number of temperatures.
For $T<T_c$ the curve
starts at a certain minimum value of the magnitude $\phi_0$
and these starting points are connected by the dotted curve,
while the dashed curve shows the result
obtained when the temperature is neglected.
The arrows point to the minima of $F_T$;
for $H>0$ there is only a single minimum,
while for $H=0$ the degenerate minima form the surface of a 4-sphere.
For each temperature the location of the minima are indicated by the solid dots.
(For $H=0$ the shallow secondary minima at $\phi_0=0$ are shown by open dots.)
\efig
\bfig   
\vspace{1in}\hspace{-0.5in}
\rotateright{\psfig{figure=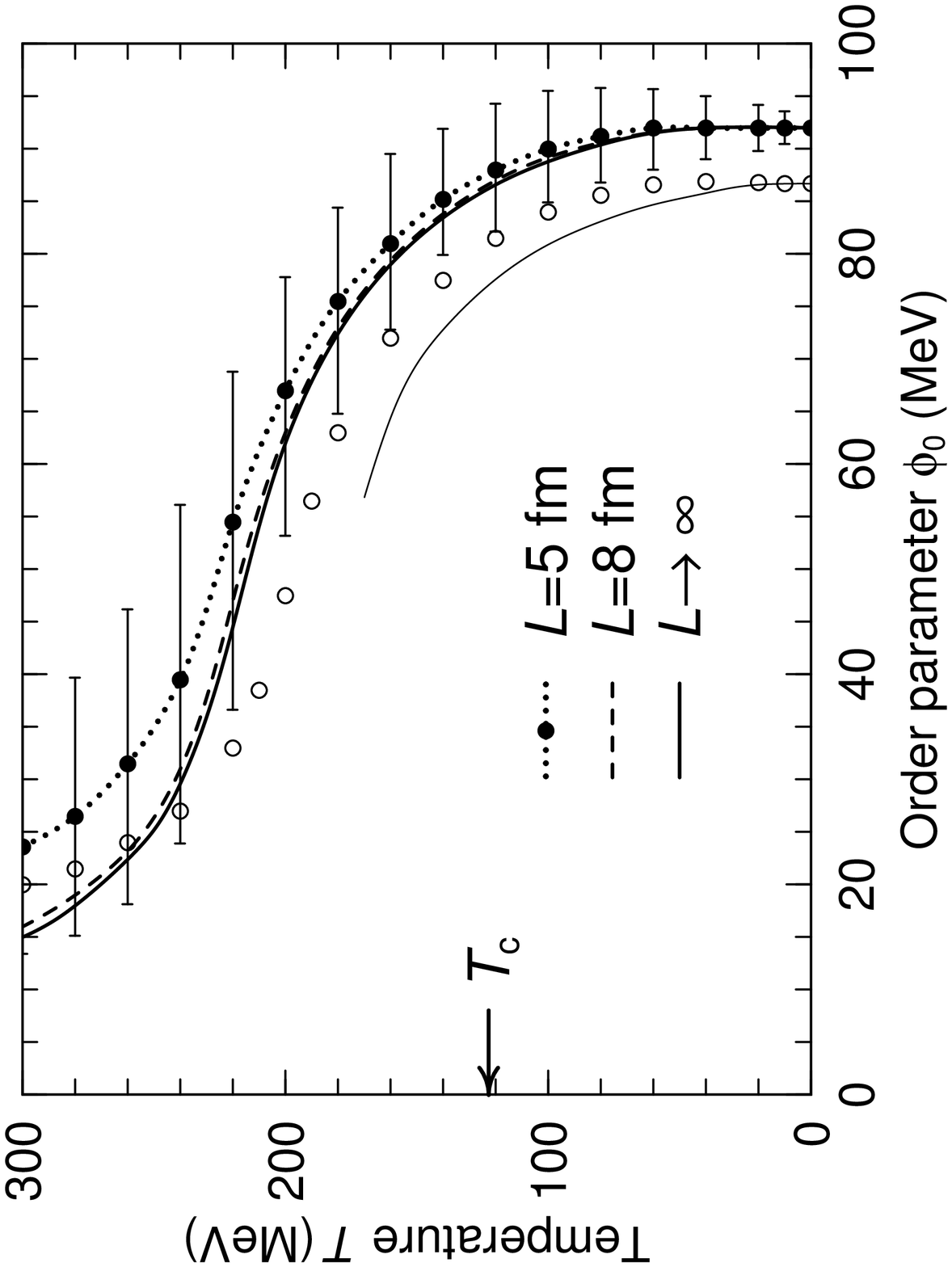,width=6in,height=4.8in}}
\vspace{-1in}
\caption{Temperature dependence of the order parameter.}
\label{f:Tphi}
The most probable value of $\phi_0$, the magnitude of the order parameter,
in the standard model where $H>0$.
In the limit of a very large box, $L\to\infty$ (solid curve)
$\phi_0$ is constrained to the value for which the free energy density
has its minimum (see fig.\ \protect\ref{f:F}$b$).
The bars show the full width at half maximum
of the thermal distribution of $\phi_0$ in the system with $L=5\ \fm$;
those for $L=8\ \fm$ are about half that size (see fig.\ \protect\ref{f:phi}).
The open dots show the centroids for $H=0$ for the box with $L=5\ \fm$,
while the light curve shows the corresponding result for a large box. 
\efig
\bfig\vspace{0.3in}
\vspace{1in}\hspace{-0.5in}
\rotateright{\psfig{figure=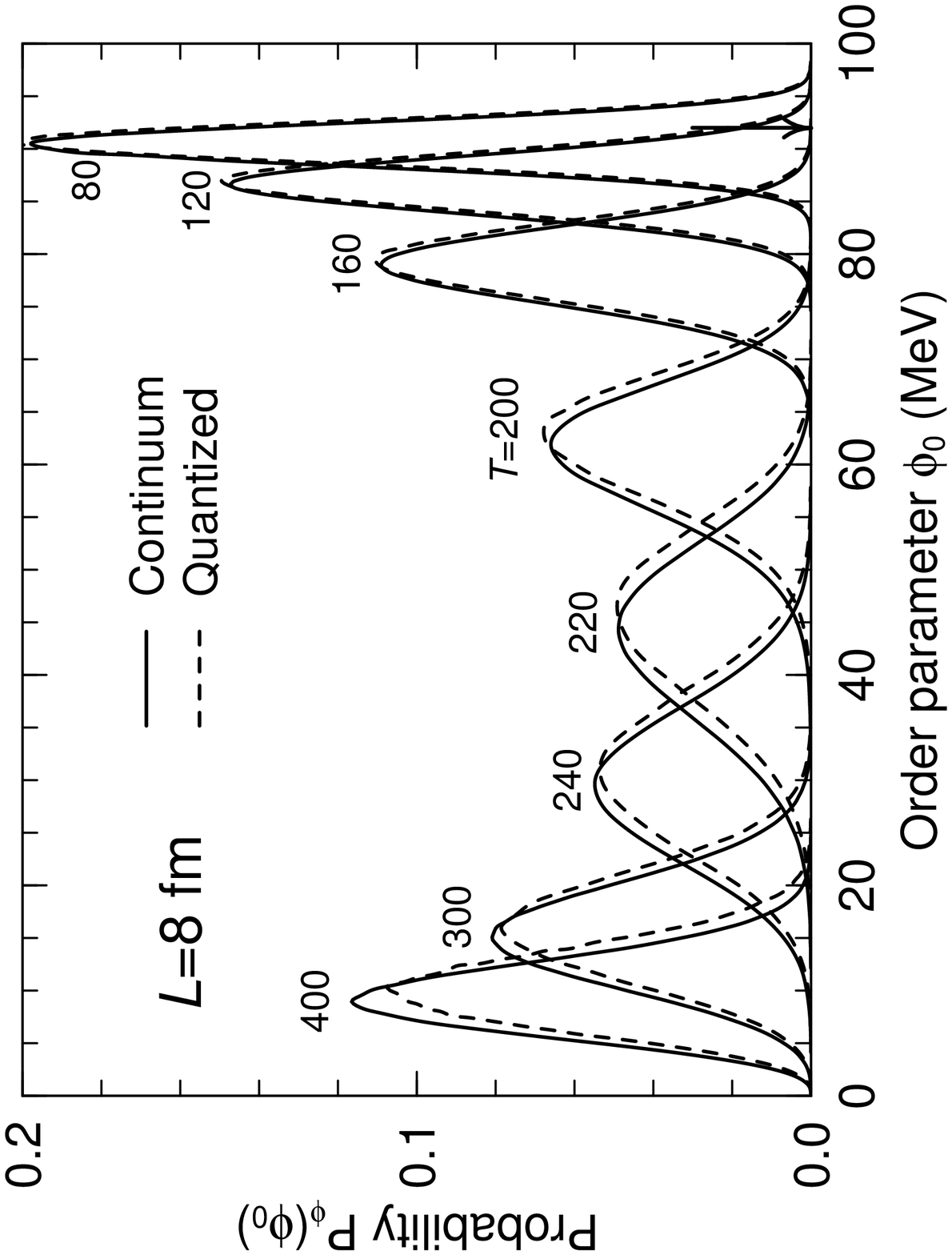,width=6in,height=4.8in}}
\vspace{-1in}
\caption{Distribution of the magnitude of the order parameter.}
\label{f:phi}
For a cubic box of side length $L=8\ \fm$ is shown
the probability density for the magnitude of the order parameter, $P(\phi_0)$,
for a range of specified temperatures $T$ (indicated),
obtained either by scaling the continuum result (solid curves)
or by quantizing the quasi-particle modes (dashed curves).
\efig
\bfig   
\vspace{1in}\hspace{-0.5in}
\rotateright{\psfig{figure=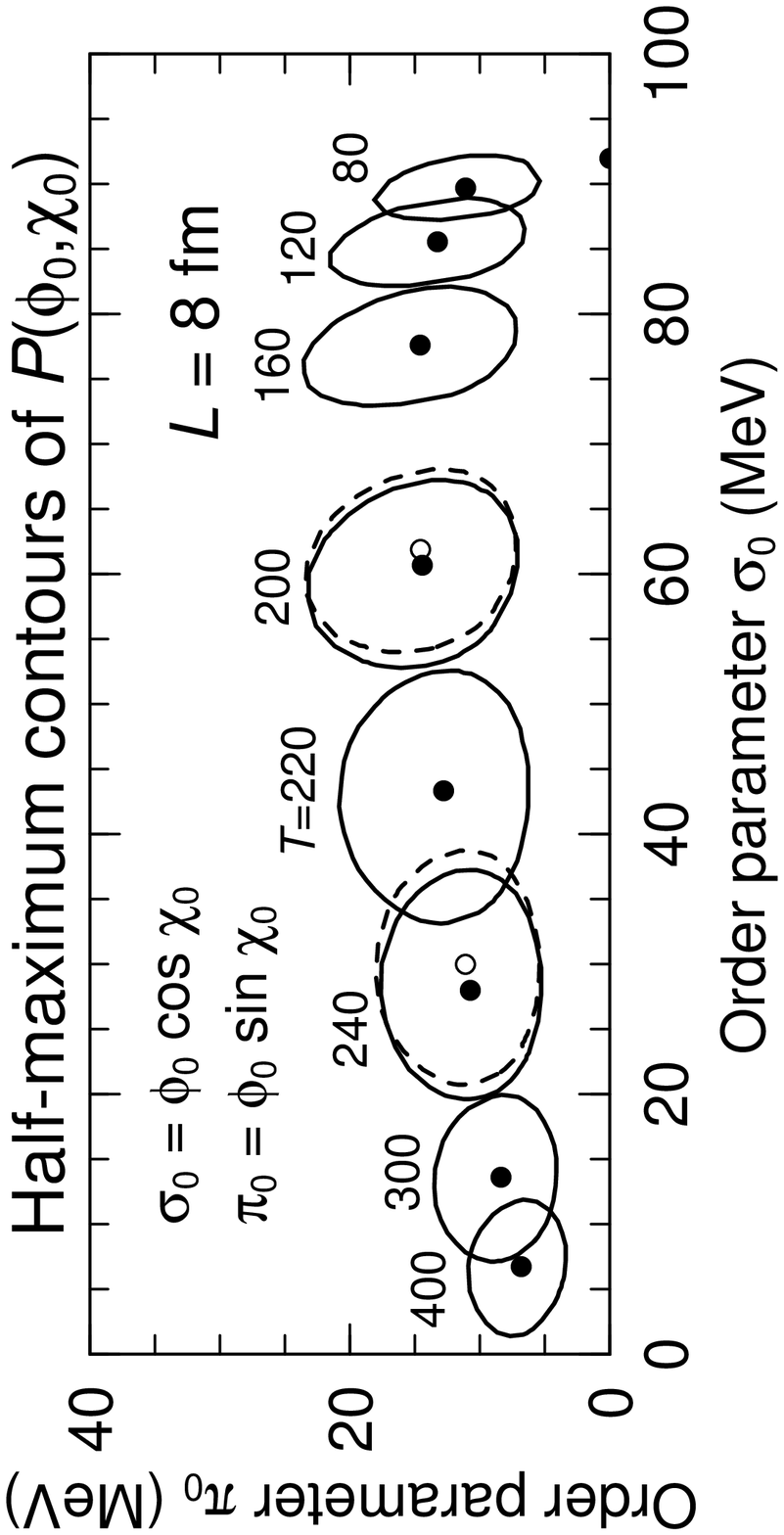,width=6in,height=4.8in}}
\vspace{-1in}
\caption{The joint distribution $P(\phi_0,\chi_0)$.}
\label{f:contour}
The projection of the probability density $P(\ul{\pphi})$
onto the variables $\phi_0$ (the magnitude of the order parameter)
and $\chi_0$ (the disalignment angle)
is displayed as a function of $\sigma_0=\phi_0\cos\chi_0$
and $\pi_0=\phi_0\sin\chi_0$,
for a cubic box of side length $L=8\ \fm$.
For each temperature $T$,
the solid dot indicates the location of the maximum of $P(\phi_0,\chi_0)$
and the solid curve traces out the half-maximum contour,
obtained by scaling the continuum results.
For the temperatures 200 and 240 meV is indicated the corresponding result
of quantizing the quasi-particle modes (dashed contours and open centroid dots).
\efig
\bfig
\vspace{1in}\hspace{-0.5in}
\rotateright{\psfig{figure=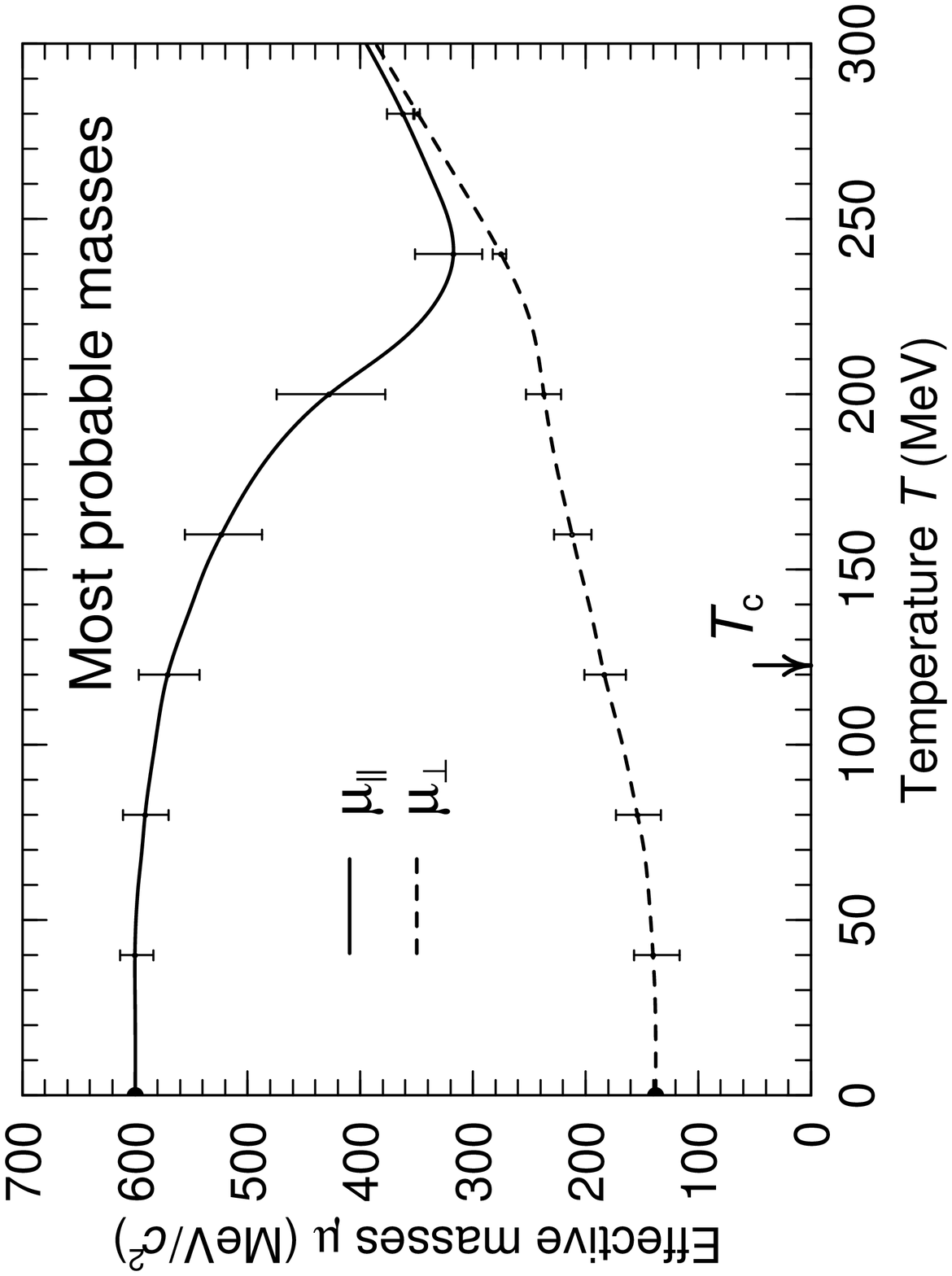,width=6in,height=4.8in}}
\vspace{-1in}
\caption{Most probable effective masses.}
\label{f:Tmu}
The most probable effective masses $\mup$ (solid) and $\mut$ (dashed)
are shown as functions of the temperature $T$,
with the error bars indicating the full width at half maximum
of the distribution resulting for a box with a side length of $L=8\ \fm$.
The results obtained by scaling the continuum results
are indistinguishable from those of the corresponding quantized treatment.
\efig
\bfig\vspace{0.4in}
\vspace{1in}\hspace{-0.5in}
\rotateright{\psfig{figure=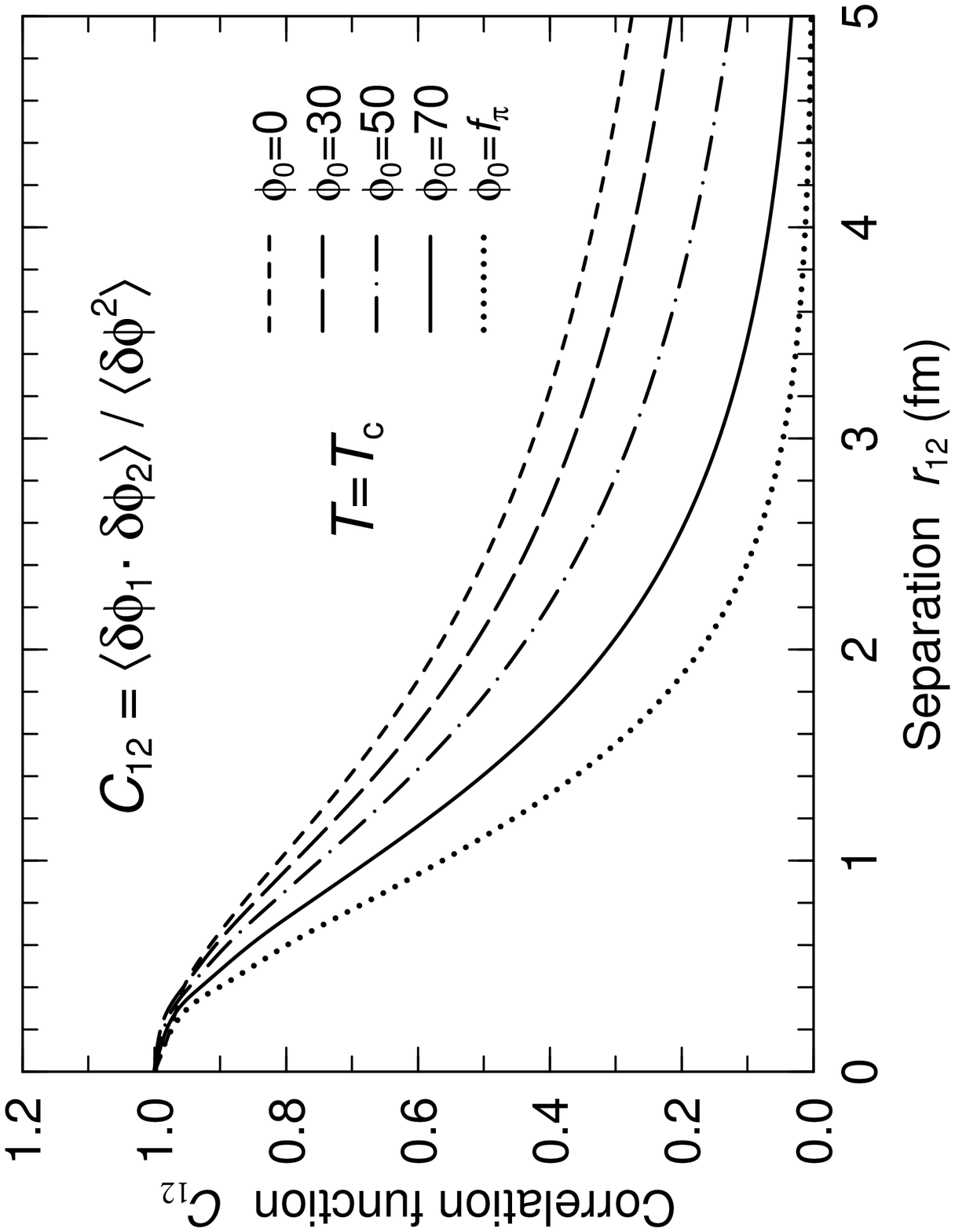,width=6in,height=4.8in}}
\vspace{-1in}
\caption{Correlation function at $T=T_c$.}
\label{f:C12-Tc}
The reduced correlation function
$C_{12}=\prec\delta\pphi(\r_1)\cdot\delta\pphi(\r_2)\succ
/\prec \delta\phi^2\succ$
calculated at the critical temperature
for various magnitudes of the order parameter,
ranging from $\phi_0=0$ to the vacuum value,
$\phi_{\rm vac}=f_\pi=92\ \MeV$.
The most probable value at $T=T_c$ is $\phi_0\approx87\ \MeV$.
\efig 
\bfig   
\vspace{1in}\hspace{-0.5in}
\rotateright{\psfig{figure=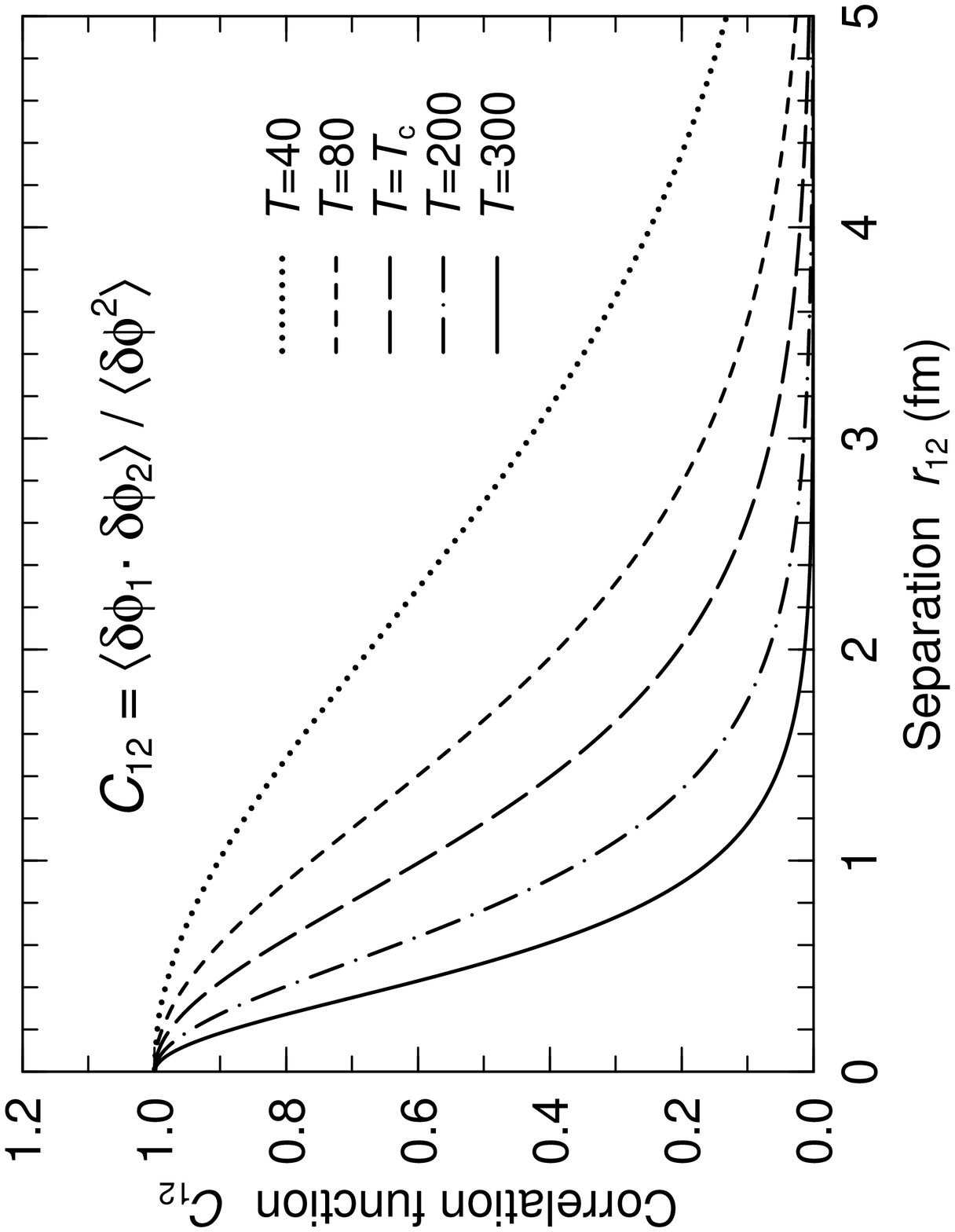,width=6in,height=4.8in}}
\vspace{-1in}
\caption{Temperature dependence of correlation function.}
\label{f:C12}
The reduced correlation function $C_{12}$ for a range of temperatures $T$,
employing for each one the most probable value of $\phi_0$,
the magnitude of the order parameter.
\efig 
\bfig   
\vspace{1in}\hspace{-0.5in}
\rotateright{\psfig{figure=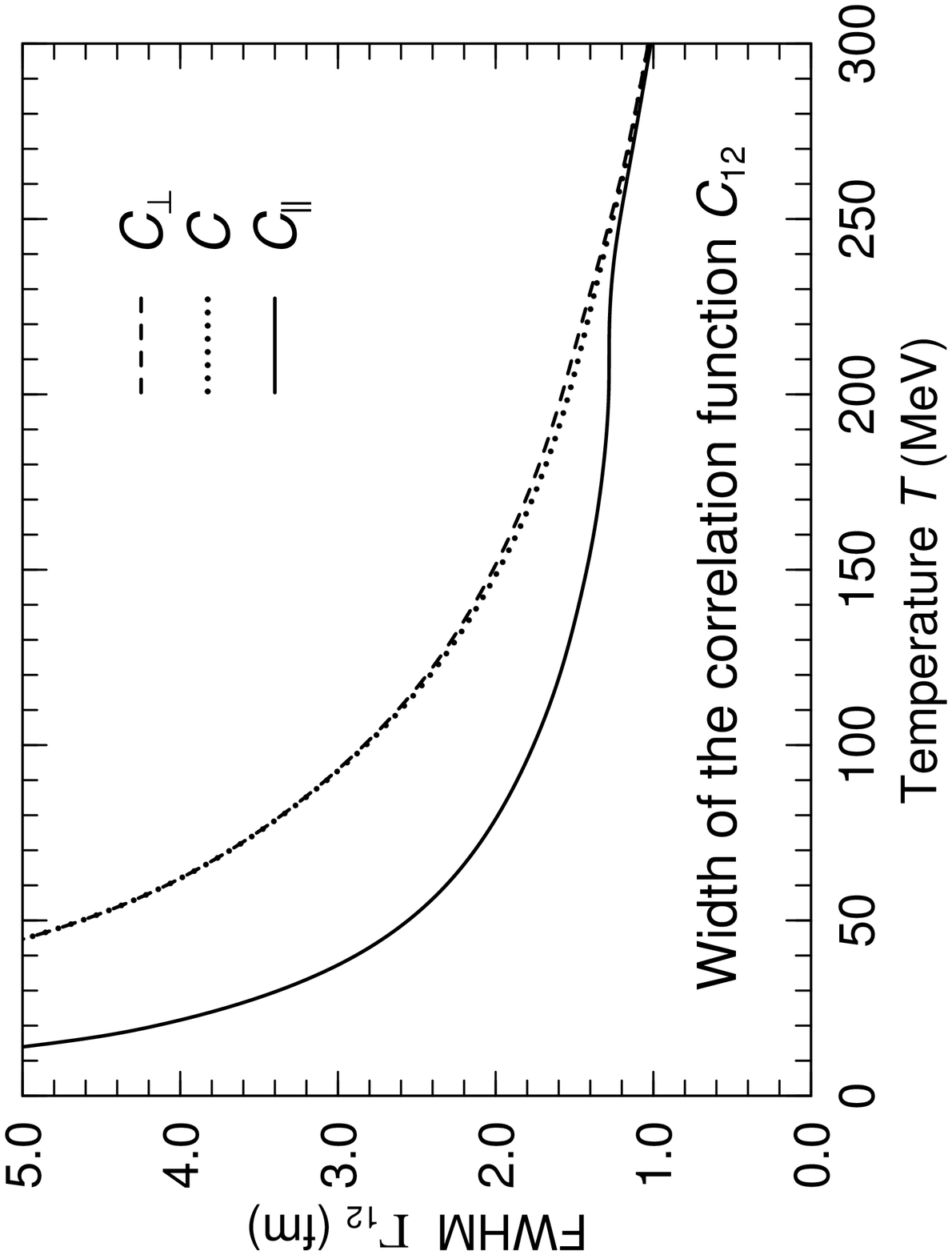,width=6in,height=4.8in}}
\vspace{-1in}
\caption{Correlation length.}
\label{f:FWHM}
The correlation length $\Gamma_{12}$
(the full width of $C_{12}$ at half maximum)
as a function of temperature,
employing the most probable value of $\phi_0$ for each $T$.
\efig 
\bfig   
\vspace{1in}\hspace{-0.5in}
\rotateright{\psfig{figure=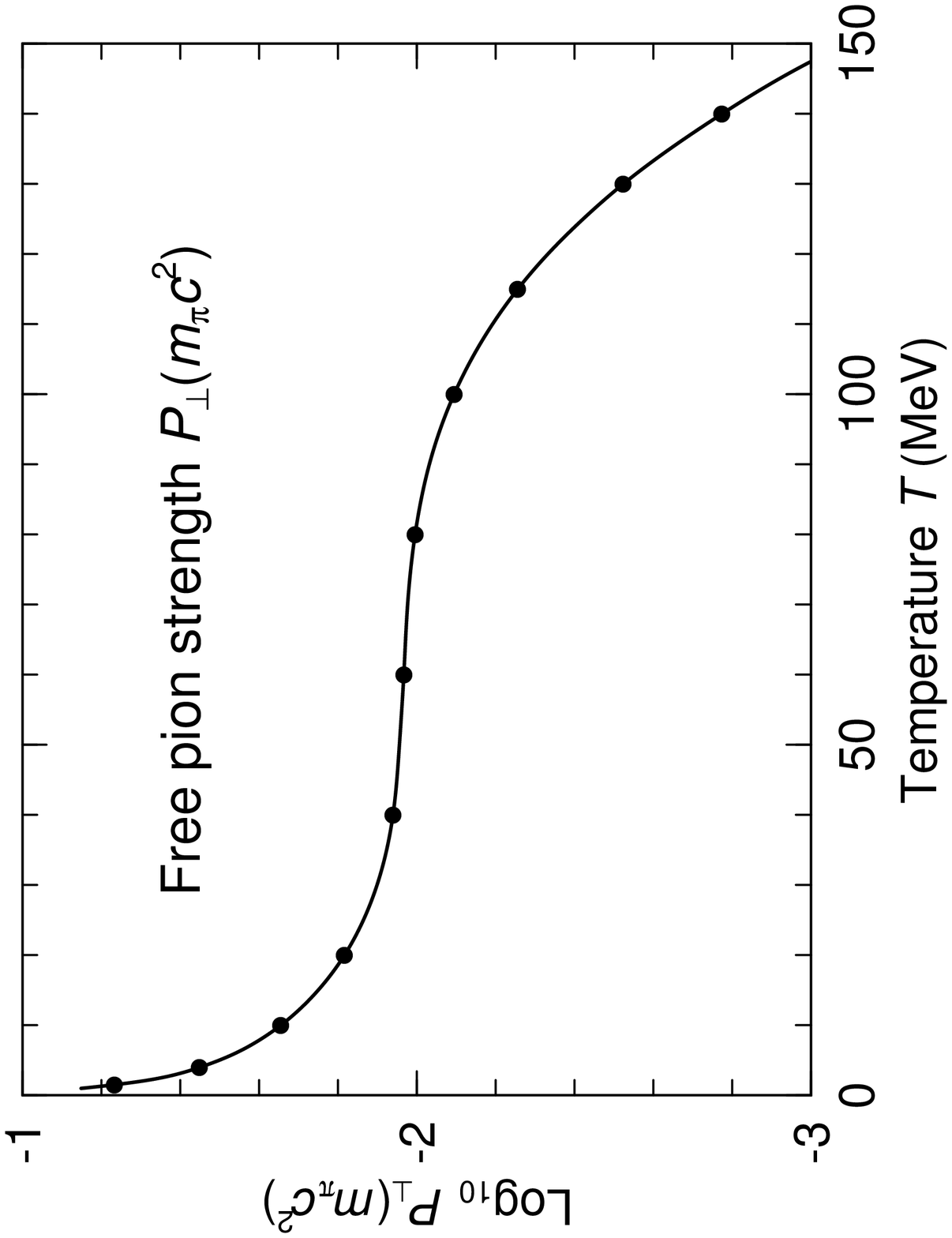,width=6in,height=4.8in}}
\vspace{-1in}
\caption{Pion production rate.}
\label{f:pions}
The strength function $P_\perp(m_\pi c^2)$
determining the overall rate at which $\pi^0$ mesons are being produced
(see eq.\ (\protect\ref{nu0})) in a source in thermal equilibrium,
as a function of its temperature $T$.
This result is obtained by performing a Fourier transform
of the quasi-particle correlation function
associated with a given order parameter $\phi_0$
and subsequently averaging over its distribution, $P_0(\phi_0)$.
The overall normalization is arbitraty.
\efig
\bfig
\vspace{1in}\hspace{-0.5in}
\rotateright{\psfig{figure=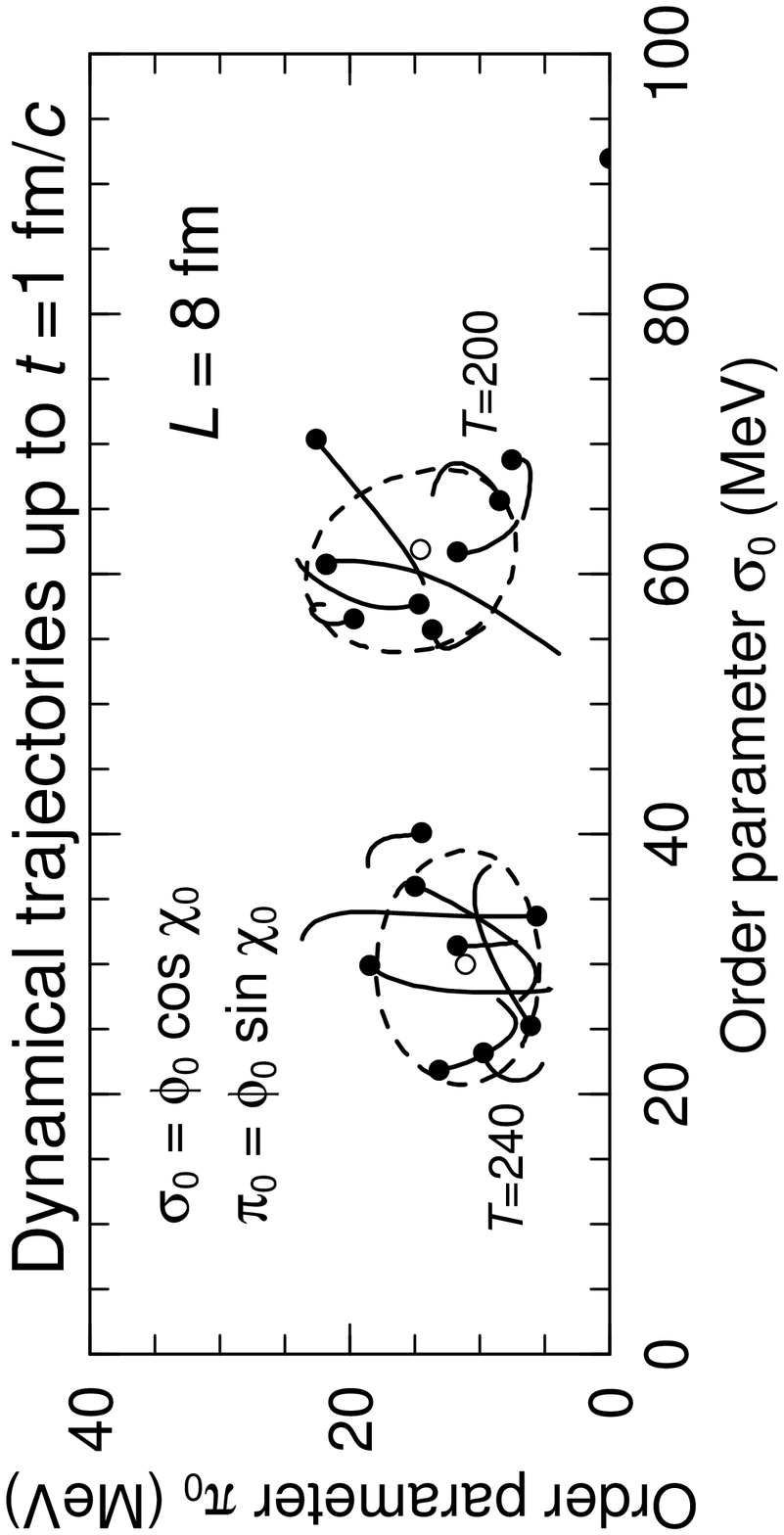,width=6in,height=4.8in}}
\vspace{-1in}
\caption{Early dynamics of the order parameter.}
\label{f:dyn0}
The early trajectories of the order parameter is shown
for a sample of eight configurations,
considering a box with $L=8\ \fm$
with a temperature of either 200 or 240 MeV.
The display is similar to that in fig.\ \protect\ref{f:contour}
and the half-density contours (dashed) 
as well as the centroids (open dots) are those already given there.
For each individual trajectory,
the initial location is indicated by the solid dot
and the attached solid curve traces out the dynamical path
up to the time $t=1\ \fm/c$.
\efig 
\bfig
\vspace{1in}\hspace{-0.5in}
\rotateright{\psfig{figure=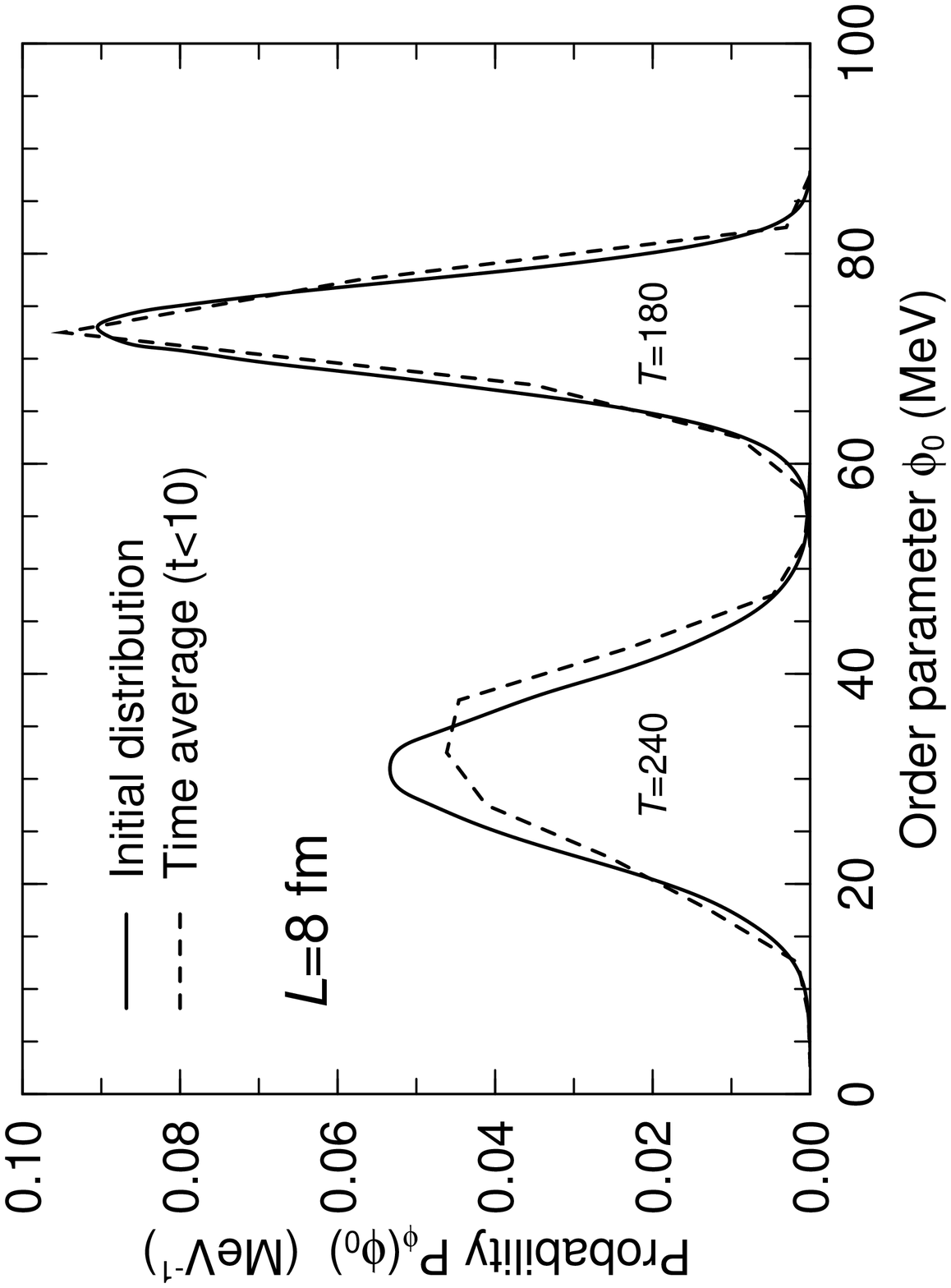,width=6in,height=4.8in}}
\vspace{-1in}
\caption{Time-averaged distribution of the order parameter.}
\label{f:dynphi0}
This figure illustrates the influence of the time evolution
on the distribution of the magnitude of the order parameter, $\phi_0$,
for a bow with $L=8\ \fm$ and for the temperatures of 180 and 240 MeV.
The solid curves show the initial distribution of $\phi_0$,
as given by approximate statistical distribution $P_\phi(\phi_0)$
(see eq. (\protect\ref{Pphi})).
A sample of 40 systems are then followed up to the time $t=10\ \fm/c$
and the order parameter is binned at regular time intervals throughout
the evolution,
leading to the dashed curves.
\efig
\bfig
\vspace{1in}\hspace{-0.5in}
\rotateright{\psfig{figure=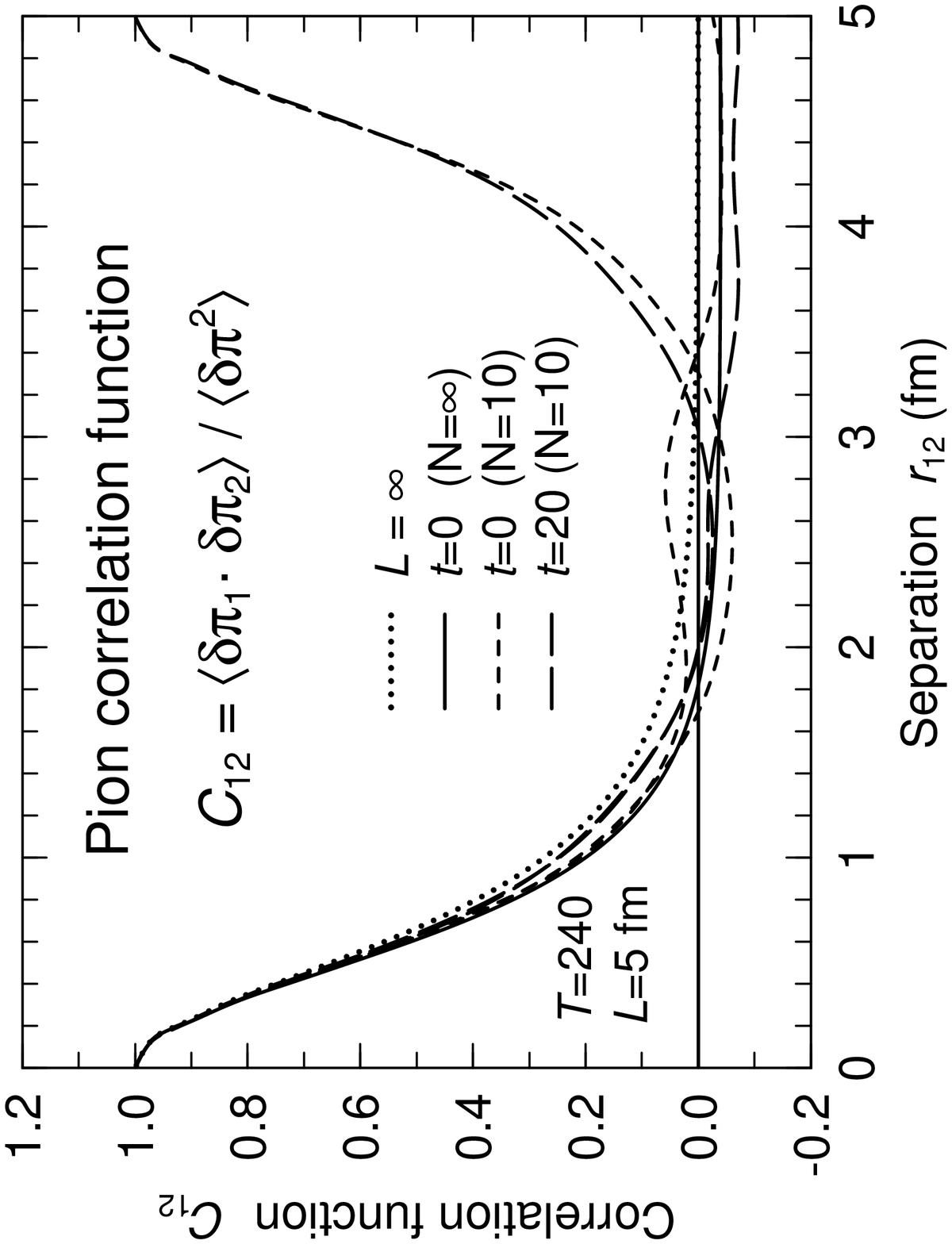,width=6in,height=4.8in}}
\vspace{-1in}
\caption{Time evolution of the correlation function.}
\label{f:dynC}
The pion correlation function $C_{12}^\pi(s_{12})$
at the temperature $T=240\ \MeV$.
The dotted curve is the continuum limit ($L\to\infty$)
and the solid curve is the corresponding thermal result
for a quantized finit box with a side length of $L=5\ \fm$.
The correlation function for a sample of ten initial configurations
are shown by the short-dashed curves,
and the long-dashed curves show the corresponding result
after they have been propagated self-consistently
up to the time $t=10\ \fm/c$.
The dashed curves have been obtained in two different ways:
The curves that go up again result from aligning the separation $\r_{12}$
along one of the cartesian directions,
while the other two are obtained for separations directed along a diagonal.
The aligned curves have a periodicity equal to $L$,
whereas the periodicity of the diagonal curves is $\sqrt{3}$ times larger.
\efig 
                        \end{document}